\documentstyle[prl,aps,axodraw,floats]{revtex}

\renewcommand{\theequation}{\thesection.\arabic{equation}}
\renewcommand{\thefootnote}{\arabic{footnote}}
\newcommand\ZZ{\hbox{\zfont Z\kern-.4emZ}}
\font\zfont = cmss10 

\def\bi{\bibitem}

\newcommand{\para}{_\parallel}
\newcommand{\pr}{_\perp}

\begin{document}

\begin{center}
{\huge{\bf Pseudo-scalar Photon Mixing In A Magnetized Medium.}} \\

\renewcommand{\thefootnote}{\fnsymbol{footnote}}
{\sf Avijit K. Ganguly $^{\!\!}$
\footnote{E-mail address: avijitk@hotmail.com}}
\vskip 10pt
{\small \it Haldia Institute Of Technology,\\
Indian Center for Advancement Of Research and Education Complex.\\
Haldia 721657,
India.}\\
\end{center}

\vglue 0.3truecm
\begin{abstract}
\begin{center}
{\bf{Abstract }}\\
Axions are pseudo-scalar particles, those arise because of breaking of
Peccei Queen (PQ) symmetry. Axions have a tree level coupling to two 
photons. As a consequence there exists a tree level coupling of axion
to photon in a magnetic field. However, in an external magnetic field, 
there exists a new loop induced, axion  photon vertex, that gives rise to 
axion photon coupling. The strength of the tree level axion photon coupling 
in magnetic field is known to be model dependent. 
However in a magnetic field, the new loop induced coupling has some interesting 
features. This note discusses the new axion photon vertex in a magnetized medium 
and the corrections arising from there. The magnitude of the correction to 
axion photon coupling, because of magnetized vacuum and  matter is estimated in 
this note. While making this estimate we note that the form of the 
axion photon vertex is related to the axial polarization tensor. This 
vertex  is shown to satisfy the Ward identity. The coupling is shown to 
have a momentum dependent piece in it.  Astrophysical importance 
of this extra modification is also pointed out.
\end{center}
\noindent
\end{abstract}


\renewcommand{\thefootnote}{(\arabic{footnote})}

\section{Introduction}
\label{sec:intro}
\setcounter{equation}{0}
\setcounter{footnote}{0}

%
%
\noindent
Axions play an important role in the conceptual aspects of particle physics
today.
They are believed to be associated with  spontaneous breaking of global Chiral symmetry $U(1)_{PQ}$ (Peccei Queen symmetry), postulated to provide an elegant solution to strong CP problem ~\cite{Peccei77,WW}. 
They are ultralight pseudo-scalar  field \cite{Peccei96}. In the
 Weinberg-Wilczek-Peccei-Queen model ( original ), the symmetry breaking scale 
was assumed to be around weak scale, $f_w$. Although the original model, 
associated with the spontaneous breakdown
of the global PQ symmetry at the Electro Weak scale (EW) $f_w$,
is excluded experimentally, modified versions of
the same with their associated axions are still of interest;
where the symmetry breaking scale is assumed to lie between the 
EW scale and $~10^{12}$ GeV.
Since the breaking scale of the PQ symmetry, $f_a$, is much larger than
the electroweak scale $f_a \gg f_w$, the resulting 
axion turns out to be  very weakly interacting (coupling constant 
$\sim f_a^{-1}$), very light ($m_a \sim f_a^{-1}$) and is often
 called ``the invisible axion model''.  

%

\noindent
Till date, it remains elusive to 
experimental confirmation, however  there have been some  efforts to 
constrain  its parameters  through various cosmological or 
astrophysical considerations. 
For instance, cosmological observational constraints  
put  bounds on its mass (such that the universe is not over closed).
Through such arguments, the allowed range for the axion mass $m_a$
has turned out to be
~\cite{Turner,Raffelt90,Raffelt-castle97,Raffelt-school97},
\begin{equation} 
10^{-5} \, \mbox{eV} \lesssim m_a \lesssim  10^{-2} \, \mbox{eV}. 
\label{eq:MsAx}
\end{equation}
\noindent
Apart from the one mentioned above, there are astrophysical
considerations too that constrain axion coupling to photons or
fermions. For instance, if they exist,  being very weakly
interacting particle, axions can drain away 
energy from stellar interiors. Since they are produced through 
processes like, $~~e^+ + e^- \to \gamma + a~~$ or 
the cross channel reaction, $~~e^-  + \gamma \to e^- + a ~~$ or 
$~~\gamma_{plasmon} \to \gamma + a~~$, $\gamma + \gamma \to a$ etc.; 
for a given coupling constant and mass,  one can estimate 
the rate at which the axions would draw away energy form
the stars. The bounds are placed by fact that,
for a given energy budget of a star, the amount of
energy drained by axion emission should be less than
it's observed luminosity.
Apart from the ones mentioned above, there also have been
experimental search for solar axions, through
the conversion of an axion into a photon in a cavity, in presence
of an external magnetic field. These searches also have placed some
bound on the axion photon coupling.
Incidentally its worth  noting that
since most of the astrophysical objects are associated with magnetic
field the same process (or the reverse of it) can take place even
in astrophysical environments too.\\

\noindent 
The coupling of axion to photon is realized through a term in the 
Lagrangian of the following form,
\begin{equation}
{\cal L} = \frac{1}{M \,}\, a \,\mathbf{E}\cdot\mathbf{\cal B}.
\label{axlagrangian}
\end{equation}
Where $a$ is the axion and $M$ is the axion coupling mass scale.  
The experimental bound on $M$ coming from the study of  solar 
axions is set to be, $M > 1.7 \times 10^{9 - 11}GeV$~\cite{Moriyama}.
It may be worth pointing out here that, though it is usually
believed that, $M > 1.7 \times 10^{9 -11}GeV$, but this is a model
dependent number. It should be noted here that the PQ symmetry breaking scale
parameter $f_a$ is proportional to $M$.
A detailed survey of various astrophysical
bounds on  the parameters of axion models and constraints on them, 
can be found in Ref.~\cite{Raffelt-book}.\\

\noindent
Since most of the bounds on axion parameters arise from astrophysical
and cosmological studies where medium  and a magnetic field are 
present, it becomes important to seek the modification of   the axion 
coupling to photon, in presence of a medium or magnetic field or both.
Particularly in some astrophysical situations where the magnetic component, 
along with medium (usually referred as magnetized medium) dominates.
Examples being, the Active Galactic Nuclei (AGN), Quasars,  Supernova,
 the Coalescing Neutron Stars or  Nascent Neutron Stars, etc., to name a
few. 
Apart from the ones discussed before, lately there seem to
be some observational signature for possible existence of astrophysical 
objects,( called Magnetars ), with magnetic field strength, ${\cal B} \sim 
10^{15} - 10^{17}$~G, i.e.,  significantly above the critical, Schwinger 
value ${\cal B}_e = m^2_e / e \simeq 4.41 \times 10^{13}$~G
~\cite{toroidal,poloidal}. However we would like to emphasize
that even normal astrophysical objects are always associated with
magnetic field, though the strength of the field may not be as strong as
${\cal B}_e$. Thus justifying the role of magnetized medium on axion
properties. Also recently there has been an attempt to describe
the observed faintness of Type Ia Supernova remnants, based on
axion photon oscillation in the magnetic field of the intergalactic
medium.\\ 

\noindent
In view of these interesting physical applications of axion physics
in astrophysics as well as cosmology it seems timely to
find out the effect of medium and magnetic field, on the couplings
of axions to photons. \\

\noindent
In this note we would investigate the matter induced
photon axion coupling in a magnetized medium. Where the particle in the
plasma will be considered to be mostly electrons, though it could be any 
fermion. Since the temperature in these astrophysical objects are not too
large (of the order of hundred MeV or so at the most) this seems to be a 
reasonable approximation. Of course our formulas are general enough to 
be extended to any temperature and density.\\

\indent
The organization of this document is as follows, in section II we would 
discuss  about the physics of axion photon coupling and the model 
dependent uncertainties that enter in the axion photon coupling parameter
$M$ given in eqn. [\ref{axlagrangian}]. In the same section we would also
try to give a brief over view of the existing axion models and their
type of coupling with fermions. 
In the next section (i.e. section II), following Schwinger's 
approach \cite{schw},
 we would elaborate on the details of the magnetized propagators. 
As would be discussed later,
matter induced Axion photon coupling in a magnetized medium has 
two contributions in it, one coming from the magnetized vacuum and the
other from the magnetized medium. 
Sections III and IV  would deal with the details of those contributions. \
Finally at the end we would conclude by justifying our results through a
general analysis and  order of magnitude estimation of the modifications
one is getting from the presence of magnetized plasma. Finally we  would  
like to conclude by pointing out the possible applications of our result.
%
\section{Axion models.}
%
After the brief motivation of  section [\ref{sec:intro}], 
we would like to review the relevant details for the popular axion
models, since they would be useful  for the medium induced
modifications to axion photon coupling. Usually there are three types 
of invisible axion models found in the literature. 
(a)the Dine-Fischler-Srednicki-Zhitnitskii, usually referred as  DFSZ 
\cite{DFSZ} model, (b) The Kim-Shifman-Vainshtein-Zakharov (KSVZ) model
\cite{KSVZ} and lastly (c) the variant invisible axion model (VIA) 
\cite{VIA}. The DFSZ axion has two doublets, $\phi_{i}(i=1,2)$ and one 
singlet $\chi$ Higgs fields; the KSVZ model contains one $\phi$ and 
one $\chi$ Higgs fields along with a super heavy exotic quark that is 
singlet under SU(2)xU(1). The  VIA  model has some similarity to that of
DFSZ model, except for the fact that it carries an extra Higgs singlet
whose phase is identified with the axion. The most widely discussed 
axion model in the literature are DFSZ and KSVZ models. Apart form
the what was mentioned at the beginning,  the main difference
between them comes out on  the basis of their tree level couplings
with leptons and quarks. The KSVZ model does not have any coupling 
to electrons at the tree level, however the same might be generated 
via radiative correction through photon-photon-axion vertex 
\cite{srednicki2}; and is higher order in coupling constant
hence not interesting.\\

The $U(1)_{PQ}$ transformation rules for the Higgs fields DFSZ model are given by
~\cite{DFSZ},
\begin{eqnarray}
\phi_{u} \to e^{i\alpha X_u} \phi_u, \mbox{\hskip 0.5cm} \phi_{d} \to e^{i\alpha X_d} \phi_d  \mbox{\hskip 0.25cm and~~}
\chi \to e^{i\alpha X_{\chi}} \chi.
\end{eqnarray}
Where $u$ and $d$ are the generation indices and $\phi_{u}$ couples
only to up type quarks and similarly $\phi_{d}$. The Higgs field 
$\chi$ interacts with $\phi_{u}$ and  $\phi_{d}$ through 
the potential term. The transformation laws for the fermions under the
same transformation are  fixed by demanding that the Yukawa interactions
would remain  invariant. In the passing it may noted that,
 following~\cite{yanagida}, one can assume for convenience that
 the left handed doublets transform like singlets under $U(1)_{PQ}$, 
thus fixing the PQ charges of the right
handed fermions through their coupling to the specifically assigned
Higgs field.

Electromagnetic coupling of axions  to photons   were   derived  by  
Kaplan~\cite{kaplan}  and  Sredeniki~\cite{sred},~\cite{bt}, using current 
algebra techniques. In order to make the PQ current color anomaly 
free, a linear combination of chiral  currents for lights quark 
are usually subtracted from the same to define the new anomaly free
axial vector current, given by,

\begin{eqnarray}
j^{a}_{\mu} = j^{PQ}_{\mu} - \frac{A^{c}_{PQ}}{1+z} \left(
\bar{u} \gamma_{\mu}\gamma_{5}u + z \bar{d} \gamma_{\mu}\gamma_{5}d
\right).
\label{current-pqa}
\end{eqnarray}
Here,  $A^{c}_{PQ}$ is the color anomaly of the PQ charge, defined
in terms  of the generators of the group $SU(3)_{\rm{c}}$ i.e.,
$\lambda_a$ (a=1 to 8) in the following way,
$~~~\delta_{ab}A^{c}_{PQ}= \rm{Tr}[ {\lambda_a \lambda_b} X_{f}]$.
Where, the trace is understood to be taken over all the Weyl fermions and 
$X_f$ are their generation specific PQ charge.\\

\noindent
%

\noindent
We note for the sake of completeness that, four divergence of 
Eqn. [\ref{current-pqa}] yields,
\begin{eqnarray}
\mbox{\hskip 1cm}\partial^{\mu}j^{a}_{\mu} = \frac{e^2}{16 \pi^2} 
 F_{\mu\nu}\tilde{F}^{\mu\nu}\left[ Tr[X_{f}Q^2_{f} ]
  - A^{c}_{PQ}\frac{2}{3}\frac{(4+z)}{(1+z)} \right] - 
\frac{A^{c}_{PQ} }{1+z} \left(
2im_u\bar{u} \gamma_{5}u + 2 z i m_d \bar{d} \gamma_{5}d
\right).
\label{current-pqa1}
\end{eqnarray}\\

\noindent
The Lagrangian describing the axion-fermion ( to be considered as lepton for the estimates
made in this note ) interaction is given by \cite{kim},    
\begin{equation}
{\cal L}_{af} = \frac{1}{f_a}j^{a}_{\mu}\partial^{\mu}a = \frac{ g'_{af}}{m_f}
\sum_{f}({\bar{\Psi}}_f \gamma_\mu \gamma_5 \Psi_f ) \, \partial^\mu \, a,
\label{eq:Ld1}
\end{equation}
where $g'_{af} = X_f m_f/f_a$ is a dimensionless
Yukawa coupling constant, $X_f$s,( the  model-dependent
factors ) are the PQ charges of different generations of quarks and leptons
 ~\cite{Raffelt-book}, as given in eqn. [\ref{current-pqa}]. 
Lastly $m_f$ is the fermion's mass.
In eqn. [\ref{eq:Ld1}] the sum over f stands for sum over all the fermions, from each family. 
Although, in places instead of using the Lagrangian given by,
[\ref{eq:Ld1}], the following form for axion fermion Lagrangian
 has also been employed,
\begin{equation}
{\cal L}_{af} =  - 2ig'_{af}\sum_f ({\bar{\Psi}}_f \gamma_5 \Psi_f) a,
\label{eq:L1}
\end{equation}
but, the correctness using [\ref{eq:Ld1}] has been pointed out
by Raffelt and Seckel in~\cite{Raffelt88}. For our calculations
we would however use [\ref{eq:Ld1}], with $g_{af}= \frac{X_f}{f_a}$.\\

\noindent
The necessity for going through the details of the model becomes more 
apparent, if one recalls the way bounds were derived for $M$, 
in eqn. [\ref{axlagrangian}]. The use of the form of that Lagrangian
 leaves much scope for the  uncertainty in estimating 
the actual PQ symmetry breaking scale --- coming from model dependent 
assignments of PQ charges of the fermion multiplets. To elaborate this,
 point we note that, the effective photon photon axion coupling is obtained
via the electromagnetic anomaly generated in the processes of making
the PQ current free of anomalous divergence coming from the strong 
interaction sector. Thus, the axion photon photon
Lagrangian as obtained from eqn. [\ref{current-pqa1}] \cite{kim},
 is given by: 
%
%
\begin{eqnarray}
{\cal L}_{a\gamma\gamma} = - \frac{e^2}{32\pi^2 f_a}\left[
A^{em}_{PQ} - A^{c}_{PQ}\frac{2(4+z)}{3(1+z)}\right] a \rm{F}{\tilde{\rm{F}}}
\label{anocoupling}.
\end{eqnarray}
%
%
In eqn. [\ref{anocoupling}], following are the definition employed,
 $A^{em}_{PQ} = \rm{Tr} [Q^2_f \cdot X_{f} ]$
and $z$ 
is  defined to be the ratio of the masses of two light quarks, i.e 
$z =\frac{ m_u}{m_d} $ and $Q_f$ is the fermion electric charge,
in units of electronic charge.
The axion photon mixing Lagrangian in an external magnetic field
turns out to be,
\begin{eqnarray}
{\mbox{\hskip 2cm}}{\cal L}_{a\gamma} = -g_{a\gamma\gamma}\frac{e^2}{32\pi^2} a \rm{F}
{\tilde{\rm{F}}}^{\rm{Ext}}, 
\mbox{~~~~~~ When~~}g_{a\gamma\gamma} =\frac{1}{f_a}\left[
A^{em}_{PQ} - A^{c}_{PQ}\frac{2(4+z)}{3(1+z)}\right] 
\label{agammab}.
\end{eqnarray}

\noindent
The mixing strength, $M$ as had already been  given in eqn.[\ref{axlagrangian}]
turns out to be, $M = \frac{32 \pi^2}{e^2 g_{a\gamma\gamma }}$. It is now rather easy to 
see that as one takes into account the effect of matter contribution
(as shown in Fig.[\ref{f:cher}]), there will be additional contributions to
$g_{a\gamma\gamma}$  given in [\ref{agammab}] and hence to $M$. Though
in principle (with out matter effects) the choice of PQ charges 
can make coupling constant $g_{a \gamma\gamma}$ extremely small, but as
we would show in this paper that as one takes into account the magnetized
matter effects, there are additional modifications to the vertex, generating
momentum dependent coupling in addition to  the details of the
model and the details of the PQ charge assignments.  We would try to establish the
same in this note.

\section{Axion photon Vertex in a Magnetized medium.}
\label{capt}
\subsection{Fermion propagator in a magnetized medium.}
\label{capt1}
We are interested in physical processes in an external
 background magnetic field.  Without any loss of
generality, the same is 
taken to be in the $z$-direction and would be denoted as ${\cal B}$. 
Any charged fermion propagator in such an external magnetic field, 
in Schwinger's approach ~\cite{schw,tsai,ditt} is given by:
\begin{eqnarray}
i S_B^V(p) = \int_0^\infty ds \, e^{\Phi(p,s)} \, G(p,s) \,,
\label{SV}
\end{eqnarray}
with $\Phi$ and $G$ defined in the following way: 
\begin{eqnarray}
\Phi(p,s) &\equiv& 
          is \left( p_\parallel^2 - {\tan (eQ_f{\cal B}s) \over eQ_f{\cal B}s} \, p\pr^2 - m^2 \right) - \epsilon |s| \,,
\label{Phi} \\
G(p,s) &\equiv& {e^{ieQ_f{\cal B}s\sigma\!_z} \over \cos(eQ_f{\cal B}s)} 
       \, \left( \rlap/p_\parallel + \frac{e^{-ieQ_f{\cal B}s\sigma_z}}
{\cos(eQ_f{\cal B}s)}\rlap/ p\pr + m \right) \nonumber \\ 
       &=& \Big[ \big( 1 + i\sigma_z \tan (eQ_f{\cal B}s) \big)
(\rlap/p_\parallel + m ) +\sec^2(eQ_f{\cal B}s) \rlap/ p\pr \Big] \,, 
\label{C}.
\end{eqnarray}
The quantity $Q_f$ stands for the charge of the respective fermions,
in units of electronic charge. Also the following useful relation should
be noted, 
\begin{eqnarray}
\sigma_z = i\gamma_1 \gamma_2 = - \gamma_0 \gamma_3 \gamma_5 \,,
\label{sigz}
\end{eqnarray}
and we have used,
\begin{eqnarray}
e^{ieQ_f{\cal B}s\sigma_z} = \cos( eQ_f{\cal B}s) + i\sigma_z \sin(eQ_f{\cal B}s) \,.
\end{eqnarray}
It should be noted that, the metric convention for this propagator
is (+,-,-,-).
To be more specific, according to the convention we follow,
\begin{eqnarray}
\rlap/ p_\parallel &=& \gamma_0 p^0 + \gamma_3 p^3 \nonumber \\                      
\rlap/p\pr &=& \gamma_1 p^1 + \gamma_2 p^2 \nonumber \\
p_\parallel^2 &=& p_0^2 - p_3^2 \nonumber \\
p\pr^2 &=& p_1^2 + p_2^2 \nonumber.
\end{eqnarray}
Of course in the range of integration indicated in Eq.~(\ref{SV}) $s$
is always positive and hence $|s|$ equals $s$ and by virtue of the 
$\epsilon$ prescription the exponent damps out at $s$ equal to infinity.
Following standard prescriptions of thermal field theory,
in the presence of a background medium, the above
propagator is modified to~\cite{elmf}:
\begin{eqnarray}
iS(p) = iS_B^V(p) + S_B^\eta(p) \,,
\label{fullprop}
\end{eqnarray}
where
\begin{eqnarray}
S_B^\eta(p) \equiv - \eta_F(p) \left[ iS_B^V(p) - i\overline S_B^V(p) \right] \,,
\end{eqnarray}
and 
\begin{eqnarray}
\overline S_B^V(p) \equiv \gamma_0 S^{V \dagger}_B(p) \gamma_0 \,,
\label{Sbar}
\end{eqnarray}
for a fermion propagator, such that
\begin{eqnarray}
S_B^\eta(p) = - \eta_F(p) \int_{-\infty}^\infty ds\; e^{\Phi(p,s)} G(p,s) \,.
\label{Seta}
\end{eqnarray}
The information about the medium is carried by, $\eta_F(p)$
that in turn carries the information about the
 distribution function for the fermions and the anti-fermions:
\begin{eqnarray}
\eta_F(p) &=& \Theta(p\cdot u) f_F(p,\mu,\beta) 
+ \Theta(-p\cdot u) f_F(-p,-\mu,\beta) \, ,
\label{eta}
\end{eqnarray}
 $f_F$ denotes the Fermi-Dirac distribution function:
\begin{eqnarray}
f_F(p,\mu,\beta) = {1\over e^{\beta(p\cdot u - \mu)} + 1} \,,
\label{distrib}
\end{eqnarray}
and $\Theta$ is the step function given by:
\begin{eqnarray}
\Theta(x) &=& 1, \; \mbox{for $x > 0$} \,, \nonumber \\
&=& 0, \; \mbox{for $x < 0$} \,. \nonumber                
\end{eqnarray}
Here the four velocity of the medium is $u$, in the rest frame it
is given by $u^{\mu}=(1,0,0,0)$. 
%
\section{ Expression for photon axion vertex in presence
 of uniform background magnetic field and  material medium.}
\begin{figure}
\begin{center}
\begin{picture}(150,40)(0,-35)
\Photon(40,0)(0,0){2}{4}
\Text(20,5)[b]{$k\leftarrow$}
\Text(75,30)[b]{$p$}
\Text(140,5)[b]{$k\leftarrow $}
\Text(47,0)[]{$\nu$}
\Text(75,-30)[t]{$p+k\equiv p'$}
\DashLine(110,0)(160,0){2}
\SetWidth{1.2}
\Oval(75,0)(25,35)(0)
\ArrowLine(74,25)(76,25)
\ArrowLine(76,-25)(74,-25)
\end{picture}
\end{center} 
\caption[]{One-loop diagram for the effective axion electromagnetic vertex .}
\label{f:cher}
\end{figure}
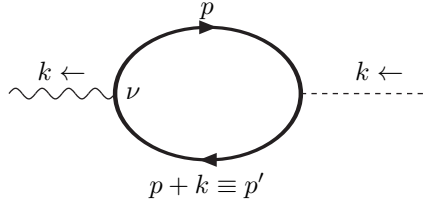

In order to estimate the loop induced $\gamma - a $ coupling, we would
start with the Lagrangian given by Eqn. [\ref{eq:Ld1}]. The effective
vertex for the $\gamma - a$ coupling can be written as ( for the sake of 
brevity, we define the notation, $p'=p+k$):
\begin{eqnarray}
\Gamma_{\nu}(k) = (-i\,g_{af}\,e\,Q_f)k^{\mu}\!\!\!\int \!\!{{d^4 p}\over{(2\pi)^4}}
\mbox{Tr}
\left[ \gamma_\mu \gamma_5 iS(p) \gamma_\nu iS(p')\right].
\label{eq:v1}
\end{eqnarray}
The effective vertex given by [\ref{eq:v1}], is computed from the diagram 
given 
in [Fig.\ref{f:cher}]. One can easily recognize that, eqn. [\ref{eq:v1}], 
has the 
following structure, $\Gamma_{\nu}(k)= k^{\mu}\Pi^A_{\mu \nu}(k)$.
Where $\Pi^A_{\mu \nu}$, is the axial polarization tensor,  comes from the 
axial coupling of the axions to the leptons and it is:
\begin{eqnarray}
i\Pi^A_{\mu \nu}(k) \!\!=(-1)(-i)^2(g_{af}\, e\,Q_f) \!\!\!\int\!\! {{d^4 p}\over {(2\pi)^4}}\mbox{Tr}\left
[ \gamma_\mu \gamma_5 iS(p) \gamma_\nu iS(p')\right].
\label{pi5}
\end{eqnarray}
The axial polarization tensor, $\Pi^A_{\mu \nu}$, would in general have
contributions from pure magnetic field background, as well as magnetic 
field plus medium (in the text, the same might as well be referred as 
magnetized medium). The Pure magnetic field contribution (i.e 
the contribution devoid of any thermal phase space factors) and the one with 
magnetized medium effects, are to be found in the following expression,
\begin{eqnarray}
\,\,i\Pi^{A}_{\mu\nu}\!(k)\!=\!(-1)(-i)^2( g_{af}\,e\,Q_f)\!\!\int\!\! {{d^4 p}\over {(2\pi)^4}}\mbox{Tr}\left
[\gamma_\mu \gamma_5 iS^V_B(p) \gamma_\nu iS^V_B(p')
\right.
&+& 
\left. 
\gamma_\mu \gamma_5 S^\eta_B(p) \gamma_\nu
iS^V_B(p')
\right.\nonumber\\
&
+
&\left. 
\gamma_\mu
\gamma_5 iS^V_B(p) \gamma_\nu S^\eta_B(p')\right].
\label{pi-ini}
\end{eqnarray}
The pure magnetic field contribution has already been estimated
in {\cite{gal,soko,hari,iorf,schu}}; however we have rechecked
and verified the same according to our convention and would report 
about it in the next section. Following that, the thermal part would 
be the one, we would be dealing with. Incidentally since we are dealing 
with the dispersive effects, the contributions coming from the absorptive 
part of the diagram are ignored ( i.e we have ignored the appearance of the 
terms in  $Tr\left[\gamma_{\mu}\gamma_5S^{\eta}(p)\gamma_{\nu}S^{\eta}(p')\right]$ 
in side the loop integral ). This is justified further by the fact that
we are interested in photons far below the threshold of pair production.
%
%
\subsection{Contribution From Magnetized Vacuum.}
%
The VA response function in a magnetic field $\Pi^{A_B}$ has been calculated in
Ref.~\cite{gal,soko,hari,iorf,schu}. However since our conventions are
different; also since there are some discrepancies
in the results stated there ( mostly  typographical though ),
we have redone the calculations and our result reads, 
\begin{eqnarray}
\label{p5n}
\Pi^{A_B}_{\mu \nu}(k)\! = \!
\frac{ig_{af} (e \,Q_f)^2}{(4\pi)^2}\!\!
\int_0^{\infty}\!\!\!\!dt \!\!\!\int_{-1}^{+1}\!\!  dv 
\,e^{\phi_0} 
\Bigl\{\Bigl(\frac{1-v^2}{2}k_{\|}^2 -2m^2_e &&\Bigr)\widetilde{F}_{\mu \nu}
- \! (1-v^2)k_{\mu_\|}(\widetilde{F} k)_{\nu}\nonumber\\
&&\hskip1em+\,
R\Bigl[k_{\nu_\bot}(k\tilde{F})_{\mu}
+k_{\mu_\bot} (k\widetilde{F})_{\nu}\Bigr]\Bigr\},
\end{eqnarray}

\begin{eqnarray}\label{Rdef}
\!\mbox{Where,}\,R=\!\!\left[\frac{1-v \sin {\cal{Z}}v \sin {\cal Z} - 
\cos {\cal Z} \cos {\cal Z}v}
{\sin ^2 {\cal{Z}}}\right]
\,\,\,\mbox{and} \,\,\,
\phi_0 = \!\!it \left[\frac{1-v^2}{4}k^2_{||}  -m^2 -\frac{\cos v{\cal Z} - 
\cos {\cal Z}}
{2{\cal Z} \sin {\cal Z}} k^2_{\perp}  \right] 
\end{eqnarray}
and $\widetilde{F}^{\mu \nu}=
\frac{1}{2}\epsilon^{\mu \nu \rho \sigma}F_{\rho \sigma}$
with $\epsilon^{0123}=1$ is the dual of the field-strength tensor, with
${\cal Z}= e Q_f {\cal B}t$. Therefore, following Eqn. [\ref{eq:v1}], The 
photon axion vertex in a purely magnetized vacuum, would be given by,
\begin{eqnarray}
\Gamma^{\nu}(k) &=& k^{\mu}\Pi^{A_B}_{\mu \nu}(k)
\label{eq:vacvertex}
\end{eqnarray}
The expression for $ \Gamma^{\nu}(k) $ turns out to be,
\begin{eqnarray}
\Gamma^{\nu}(k) \! &=& \!
\frac{ig_{af} (e \,Q_f)^2}{(4\pi)^2}\!\!
\int_0^{\infty}\!\!\!\!dt \!\!\!\int_{-1}^{+1}\!\!  dv 
\,e^{\phi_0} 
\Bigl\{\Bigl(\frac{1-v^2}{2}k_{\|}^2 -2m^2_e  \Bigr)\widetilde{F}_{\mu \nu}
- \! (1-v^2)k_{\mu_\|}(\widetilde{F} k)_{\nu}\nonumber\\
&&\hskip1em+\,
R\Bigl[k_{\nu_\bot}(k\tilde{F})_{\mu}
+k_{\mu_\bot} (k\widetilde{F})_{\nu}\Bigr]\Bigr\},
\label{eq:vacvertex-detailed}
\end{eqnarray}\\
\noindent
This result is not gauge invariant. However, one may integrate the
first term under the integral by parts~\cite{hari,iorf}
\begin{eqnarray}
\label{ip}
&&\int_0^\infty dt\,\Bigl(\frac{1-v^2}{2}k_{\|}^2 -2 m^2\Bigr)
e^{\phi_0}
=\,2i+\int_0^\infty dt
\,k_{\bot}^2 e^{\phi_0}R.
\label{int:byparts}
\end{eqnarray}
to arrive at,

\begin{eqnarray}
\!\!\!\!\!\! \Gamma^{\nu}(k) \! &=& \!
\frac{ig_{af} (e \,Q_f)^2}{(4\pi)^2}\!\!\Bigg[
4i (k\widetilde{F})_{\nu}+
\int_0^{\infty}\!\!\!\!dt \!\!\!\int_{-1}^{+1}\!\!  dv 
\,e^{\phi_0} 
\Bigl\{\Bigl(-(1-v^2)k_{\|}^2 \Bigr) 
(k\widetilde{F})_{\nu} \nonumber
+ R\Bigl[k_{\nu_\bot}k^{\mu}(k\tilde{F})_{\mu}
\nonumber \\ 
&+ & k^{\mu}k_{\mu_\bot} (k\widetilde{F})_{\nu}
+ k^2_{\perp}
(k\widetilde{F})_{\nu}
\Bigr]\Bigr\}\Bigg] \nonumber \\
&=&
\!
\frac{ig_{af} (e \,Q_f)^2}{(4\pi)^2}\!\!\Bigg[
4i -
\int_0^{\infty}\!\!\!\!dt \!\!\!\int_{-1}^{+1}\!\!  dv 
\,e^{\phi_0} 
\Bigl(1-v^2\Bigr)  k_{\|}^2  
\Bigg](k\widetilde{F})_{\nu}
\label{eq:vacvertex-final}
\end{eqnarray}\\
As is evident, terms proportional to R in eqn. (\ref{eq:vacvertex-final})
cancels out against each other leaving the vertex gauge invariant. 
Incidentally, in our way of writing $(k\widetilde{F})^{\nu}= k_{\mu}
\widetilde{F}^{\mu\nu}$ and $(\widetilde{F}k)^{\nu}= \widetilde{F}^{\nu\mu}k_{\mu}$. 
So the Effective Lagrangian giving rise to axion photon coupling in a 
magnetized medium would therefore be given by,
\begin{eqnarray}
{\cal L}^{B}_{a \gamma}= a A^{\nu}\Gamma_{\nu}(k)
\label{eq:vac-eff-lag}
\end{eqnarray}
It should be noted that, $a$ in Eqn.[\ref{eq:vac-eff-lag}], denotes the
axion field. 

%
%
%
%
%

In the limit of $\omega << m_f $,  axion photon vertex can be written as:

\begin{eqnarray}
{\cal L}^{\cal B}_{a\gamma} &=& \frac{-1}{32 \pi^2 } g_{ af} (eQ_f)^2 a 
F_{\mu\nu}\tilde{F}^{\mu\nu}\left[ 4+ \frac{k^2_{\perp}}{eB}
 \frac{4}{6h} \right] \nonumber \\
&=&\frac{-1}{32 \pi^2 } g_{ af} (eQ_f)^2 
\left[4 + \frac{4}{3} \left(\frac{k^2_{\para}}{m^2}\right) \right]
a F_{\mu\nu}\tilde{F}^{\mu\nu} .
\label{eq:vac-eff-lag2b}
\end{eqnarray}

Equation [\ref{eq:vac-eff-lag2b}], describes the effective axion photon
interaction.

%
%
%
\section{Contribution From the Magnetized medium}
%
%
%

Having estimated the effective axion photon vertex in a purely 
magnetic environment, we would now estimate the same in a 
magnetized medium. As before, one can do that by using the form of the 
 fermion propagator in a magnetic field in
presence of a thermal medium, as given by expressions(\ref{SV}) and 
(\ref{Seta}); on doing that, we arrive at:
\begin{eqnarray}
i\Pi^{A_{\beta  B}}_{\mu\nu}(k)&=&-(-i)^2( g_{af}e Q_f)(-1) \int {{d^4 p}\over {(2\pi)^4}}
\int_{-\infty}^\infty ds\, e^{\Phi(p,s)}\nonumber\\
& &\int_0^\infty
ds'e^{\Phi(p',s')}\big[ \mbox{Tr}\left[\gamma_\mu\gamma_5 G(p,s)
\gamma_\nu G(p',s')\right]\eta_F(p)\big. \nonumber\\
& &\big.\hskip 1cm + \mbox{Tr}\left[\gamma_\mu
\gamma_5 G(-p',s') \gamma_\nu G(-p,s)\right]\eta_F(-p)
 \big]\nonumber\\
 &=& -(-i)^2( g_{af}e Q_f)(-1) \int {{d^4 p}\over {(2\pi)^4}}
\int_{-\infty}^\infty ds\, e^{\Phi(p,s)}\nonumber\\
& & \times  \int_0^\infty
ds'\,e^{\Phi(p',s')}\mbox{R}_{\mu\nu}(p,p',s,s')
\label{compl}
\end{eqnarray}
where $\mbox{R}_{\mu\nu}(p,p',s,s')$ contains the trace part. 
It should be noted that, in general $\mbox{R}_{\mu\nu}(p,p',s,s')$
is a polynomial in powers of the external magnetic field ( with 
even and odd powers of $B$), i.e,
\begin{eqnarray}
\mbox{R}_{\mu\nu}(p,p',s,s') = \mbox{R}^{(E)}_{\mu\nu}(p,p',s,s')+ 
{\cal R}^{(O)}_{\mu\nu}(p,p',s,s')
\label{R}
\end{eqnarray}

In the sections below we would evaluate them one after another.
\subsection{$\mbox{R}_{\mu\nu}$ to even and odd orders in magnetic field}

%

We denote the pieces even and odd orders in the
external magnetic field in $\mbox{R}_{\mu\nu}$    
as  $\mbox{R}^{(E)}_{\mu\nu}$ and $\mbox{R}^{(O)}_{\mu\nu}$. 
In addition to being just even and odd in powers of $eQ_f{\cal B}$, they  also
are odd and even in powers of the chemical potential (the same would be 
clear as we go along) i.e., under charge
conjugation, ${\cal B}\& \mu \leftrightarrow (-\mu) \& (-{\cal B}) $, 
they both behave differently. More over their  parity structures are 
 also different. These properties may come very useful while analyzing,
 the structure of axion photon coupling, using discrete 
symmetry arguments; and that is the reason, why we have treated them 
separately. Calculating the traces we obtain (details
for $R^{(O)}$ is provided in the appendix, in any case $R^{E}_{\mu\nu}$
on contraction with $k^{\mu}$ vanish and if that can be established
on general grounds, one does not have to evaluate it.),
\begin{eqnarray}
& &\mbox{R}^{(E)}_{\mu\nu}=
4i\eta_{-}(p)\left[\varepsilon_{\mu \nu \alpha\para \beta\para}
p^{\alpha\para} p'^{\beta\para}(1 + \tan(eQ_f{\cal B}s)\tan(eQ_f{\cal B}s'))\right.\nonumber\\
&+&\left. \varepsilon_{\mu \nu \alpha\para
\beta\pr} p^{\alpha\para} p'^{\beta\pr} \sec^2 (e{\cal
B}s')
+\varepsilon_{\mu \nu \alpha\pr \beta\para} p^{\alpha\pr} 
p'^{\beta\para} \sec^2 (eQ_f{\cal B}s)\right.\nonumber\\
&+&\left. \varepsilon_{\mu \nu \alpha\pr 
\beta\pr} p^{\alpha\pr} p'^{\beta\pr} \sec^2 (eQ_f{\cal B}s)
\sec^2 (eQ_f{\cal B}s')\right]
\label{reven}
\end{eqnarray}
and
\begin{eqnarray}
&&{\mbox{$\cal{R}$}}^{(O)}_{\mu\nu}=4i\eta_{+}(p)\left[ - m^2\varepsilon_{\mu \nu
1 2}(\tan(eQ_f{\cal B}s) + \tan(eQ_f{\cal B}s'))\right.\nonumber\\
&+&\left.\left\{(g_{\mu \alpha\para} p^{\widetilde{\alpha\para}}
p'_{\nu\para} - g_{\mu \nu} p'_{\alpha\para}
p^{\widetilde{\alpha\para}} +g_{\nu \alpha\para} p^{\widetilde{\alpha\para}}
p'_{\mu\para} )\right.\right.\nonumber\\
&+&\left.\left.(g_{\mu \alpha\para} p^{\widetilde{\alpha\para}}
p'_{\nu\pr} + g_{\nu \alpha\para} p^{\widetilde{\alpha\para}}
p'_{\mu\pr}) \sec^2(eQ_f{\cal B}s')\right\} \tan(eQ_f{\cal
B}s)\right.\nonumber\\
&+& \left.\left\{(g_{\mu \alpha\para} p'^{\widetilde{\alpha\para}}
p_{\nu\para} - g_{\mu \nu} p_{\alpha\para}
p'^{\widetilde{\alpha\para}} +g_{\nu \alpha\para} 
p'^{\widetilde{\alpha\para}}
p_{\mu\para} )\right.\right.\nonumber\\
&+&\left.\left.(g_{\mu \alpha\para} p'^{\widetilde{\alpha\para}}
p_{\nu\pr} + g_{\nu \alpha\para} p'^{\widetilde{\alpha\para}}
p_{\mu\pr}) \sec^2(eQ_f{\cal B}s)\right\} \tan(eQ_f{\cal
B}s') \right]. 
\label{roddm}
\end{eqnarray}
We would further express Eqn.[\ref{roddm}] as a sum of two pieces, 
${\cal{R}}^{(O)}_{\mu\nu} = 4i\eta_{+}(p)
\left[ - 2 m^2\varepsilon_{\mu \nu 1 2}(\tan(eQ_f{\cal B}s) 
+ \tan(eQ_f{\cal B}s'))\right]+ \mbox{R}^{o}_{\mu\nu}(p,p',s,s')$.
Where $\mbox{R}^{o}_{\mu\nu}(p,p',s,s')$ is defined as follows,

\begin{eqnarray}
& &\mbox{R}^{(o)}_{\mu\nu}=4i\eta_{+}(p)\left[ m^2\varepsilon_{\mu \nu
1 2}(\tan(eQ_f{\cal B}s) + \tan(eQ_f{\cal B}s'))\right.\nonumber\\
&+&\left.\left\{(g_{\mu \alpha\para} p^{\widetilde{\alpha\para}}
p'_{\nu\para} - g_{\mu \nu} p'_{\alpha\para}
p^{\widetilde{\alpha\para}} +g_{\nu \alpha\para} p^{\widetilde{\alpha\para}}
p'_{\mu\para} )\right.\right.\nonumber\\
&+&\left.\left.(g_{\mu \alpha\para} p^{\widetilde{\alpha\para}}
p'_{\nu\pr} + g_{\nu \alpha\para} p^{\widetilde{\alpha\para}}
p'_{\mu\pr}) \sec^2(eQ_f{\cal B}s')\right\} \tan(eQ_f{\cal
B}s)\right.\nonumber\\
&+& \left.\left\{(g_{\mu \alpha\para} p'^{\widetilde{\alpha\para}}
p_{\nu\para} - g_{\mu \nu} p_{\alpha\para}
p'^{\widetilde{\alpha\para}} +g_{\nu \alpha\para} 
p'^{\widetilde{\alpha\para}}
p_{\mu\para} )\right.\right.\nonumber\\
&+&\left.\left.(g_{\mu \alpha\para} p'^{\widetilde{\alpha\para}}
p_{\nu\pr} + g_{\nu \alpha\para} p'^{\widetilde{\alpha\para}}
p_{\mu\pr}) \sec^2(eQ_f{\cal B}s)\right\} \tan(eQ_f{\cal
B}s') \right]. 
\label{rodd}
\end{eqnarray}
Here
\begin{eqnarray}
\eta_+(p)&=&\eta_F(p) + \eta_F(-p) \label{etaplus}\\
\eta_-(p)&=&\eta_F(p) - \eta_F(-p)
\label{etaminus}
\end{eqnarray}
they carry the informations about the thermal nature of the medium.
More over, it should be noted that,according to the convention followed 
in the text,
\begin{eqnarray}
a_{\mu_{\para}} b^{{\widetilde \mu}_{\para}}=a_0 b^3 + a_3 b^0.\nonumber
\end{eqnarray}
In the steps below it will be shown that, as Eqn.[\ref{rodd}] is contracted 
with $k^{\mu}$, it would vanish. Hence, only the non-vanishing 
contribution would be proportional to 
the integral over the four momentum and the parameters $p$, $s$ and $s'$
respectively of,

$$4i\eta_{+}(p) e^{\Phi(p,s)+\Phi(p',s')} 
\left[ - 2 m^2\varepsilon_{\mu \nu 1 2}(\tan(eQ_f{\cal B}s) 
+ \tan(eQ_f{\cal B}s'))\right] $$.\\

\noindent
In the rest frame of the medium,
$p\cdot u=p_0$. Thus the distribution function does not depend on the
spatial components of $p$. In this case we can write the expressions of
$\mbox{R}^{(e)}_{\mu \nu}$ and $\mbox{R}^{(o)}_{\mu \nu}$ using the
relations derived 
earlier {\cite{frd1}} inside the integral sign, as
\begin{eqnarray}
p^{\beta\pr}&\stackrel{\circ}{=}&-{\tan(eQ_f{\cal B}s')\over {\tan(eQ_f{\cal B}s) +
\tan(eQ_f{\cal B}s')}}\,k^{\beta\pr}
\label{pperpint}\\
p'^{\beta\pr}&\stackrel{\circ}{=}&{\tan(eQ_f{\cal B}s)
\over {\tan(eQ_f{\cal B}s) + \tan(eQ_f{\cal B}s')}}\,k^{\beta\pr}
\label{primeperpint}\\
p^2\pr&\stackrel{\circ}{=}&{1\over {\tan(eQ_f{\cal B}s) + \tan(eQ_f{\cal B}s')}}\left[-ieQ_f{\cal
B}\right.\nonumber\\
& &\hskip 1.5cm \left. + {\tan^2(eQ_f{\cal B}s')\over
{\tan(eQ_f{\cal B}s) + \tan(eQ_f{\cal B}s')}}\, k^2\pr\right]
\label{psq}\\
p'^2\pr&\stackrel{\circ}{=}&{1\over {\tan(eQ_f{\cal B}s) + \tan(eQ_f{\cal B}s')}}\left[-ieQ_f{\cal
B}\right.\nonumber\\
& &\hskip 1.5cm \left. + {\tan^2(eQ_f{\cal B}s)\over
{\tan(eQ_f{\cal B}s) + \tan(eQ_f{\cal B}s)}}\, k^2\pr\right]
\label{p'sq}\\
m^2 &\stackrel{\circ}{=}&\left(i{d\over ds} + (p^2\para -\sec^2(eQ_f{\cal B}s)
p^2\pr)\right)
\label{m:sq}
\end{eqnarray}
Coming back to $R^{E}_{\mu\nu}$, eqn.[\ref{reven}] can  further be simplified using 
Eqn.[\ref{pperpint},\ref{primeperpint}] and one arrives at,
\begin{eqnarray}
\mbox{R}^{(E)}_{\mu \nu}\stackrel{\circ}{=}&&\!\!\!\!\!\!\!\!\!
4i\eta_{-}(p_0)\left[\varepsilon_{\mu \nu \alpha\para \beta\para}
p^{\alpha\para} p'^{\beta\para}(1 + \tan(eQ_f{\cal B}s)\tan(eQ_f{\cal B}s'))
+
\varepsilon_{\mu \nu \alpha\para
\beta\pr} p^{\alpha\para} p'^{\beta\pr} \sec^2 (eQ_f{\cal
B}s')\right.\nonumber\\
&+&\left.\varepsilon_{\mu \nu \alpha\pr \beta\para} p^{\alpha\pr} 
p'^{\beta\para} \sec^2 (eQ_f{\cal B}s)\right]
\label{reven1}
\end{eqnarray}
The last term in Eqn.[\ref{reven}]has vanished  because, on using 
the identities
as mentioned above,(i.e Eqn.[\ref{pperpint},\ref{primeperpint}], it turns
out to be proportional to $\epsilon_{\mu\nu\alpha_{\perp}\beta_{\perp}}
k^{\alpha_{\perp}}k^{\beta_{\perp}}$, hence zero. 
The $\stackrel{\circ}{=}$ symbol signifies that the relations above are
not proper equations, the equality holds only inside the momentum
integrals in Eq.(\ref{compl}). 
\section{Evaluations In Even And Odd powers In Magnetic Field.}
\subsection{Evaluation of $\Pi^{A_{\beta B}}_{\mu \nu}$ to even
orders in the external field}
\label{Even-B-Field}
The  axial polarization tensor, even in powers of the external field
( denoting by $ \Pi^{A_{\beta B(E)}}_{\mu \nu} $), is given by:
\begin{eqnarray}
\Pi^{A_{\beta B(E)}}_{\mu \nu} = -(-i e)^2(-1) \int {{d^4 p}\over {(2\pi)^4}}
\int_{-\infty}^\infty\!\!\! ds\, e^{\Phi(p,s)} \int_0^\infty\!\!\!
ds'\,e^{\Phi(p',s')}\mbox{R}^{(E)}_{\mu\nu}(p,p',s,s').
\label{a1}
\end{eqnarray}
Using  Eq.(\ref{reven1}) in the rest frame of the medium, we have:
\begin{eqnarray}
\mbox{R}^{(E)}_{\mu \nu}\stackrel{\circ}{=}\!\!\!&&\!\!
4i\eta_{-}(p_0)\left[\varepsilon_{\mu \nu \alpha\para \beta\para}
p^{\alpha\para} p'^{\beta\para}(1 + \tan(eQ_f{\cal B}s)\tan(eQ_f{\cal B}s'))\right.\nonumber\\
&+&\left. \varepsilon_{\mu \nu \alpha\para
\beta\pr} p^{\alpha\para} p'^{\beta\pr} \sec^2 (e Q_f{\cal
B}s')
+\varepsilon_{\mu \nu \alpha\pr \beta\para} p^{\alpha\pr} 
p'^{\beta\para} \sec^2 (eQ_f{\cal B}s)\right].\nonumber\\
\label{a2}
\end{eqnarray}
Noting that it is possible to write, 
%
$
q^{\alpha}p_{\alpha} = q^{\alpha\para}p_{\alpha\para} + 
q^{\alpha\pr}p_{\alpha\pr}\nonumber
$
%
 eq.(\ref{a2}) can be written as,  
\begin{eqnarray}
\mbox{R}^{(E)}_{\mu \nu}\stackrel{\circ}{=}\!\!\!\!\!\!\! && \!\!
4i\eta_{-}(p_0)\left[(\varepsilon_{\mu \nu \alpha \beta}
p^{\alpha} p'^{\beta} - \varepsilon_{\mu \nu \alpha \beta\pr}
p^{\alpha} p'^{\beta\pr}
-
\varepsilon_{\mu \nu \alpha\pr \beta}
p^{\alpha\pr} p'^{\beta})(1 + \tan(eQ_f{\cal B}s)\tan(eQ_f{\cal B}s'))\right.\nonumber\\
&+&\left. \varepsilon_{\mu \nu \alpha
\beta\pr} p^{\alpha} p'^{\beta\pr} \sec^2 (eQ_f{\cal
B}s')
+\varepsilon_{\mu \nu \alpha\pr \beta} p^{\alpha\pr} 
p'^{\beta} \sec^2 (eQ_f{\cal B}s)\right].
\label{a3}
\end{eqnarray}
Here throughout we have omitted terms such as $\varepsilon_{\mu \nu
\alpha\pr \beta\pr} p^{\alpha\pr}  
p'^{\beta\pr}$, since  by the application of Eq.(\ref{pperpint}) we
have
\begin{eqnarray}
\varepsilon_{\mu \nu \alpha\pr \beta\pr} p^{\alpha\pr} 
p'^{\beta\pr} =\varepsilon_{\mu \nu \alpha\pr \beta\pr}
p^{\alpha\pr} p^{\beta\pr} + \varepsilon_{\mu \nu \alpha\pr
\beta\pr} p^{\alpha\pr}  
k^{\beta\pr} 
\stackrel{\circ}{=}\!\!\! &-&\!\! \frac{\tan(eQ_f{\cal B}s')}{\tan(eQ_f{\cal B}s')+\tan(eQ_f{\cal B}s')}
\nonumber \\ &\times&
\varepsilon_{\mu \nu \alpha\pr 
\beta\pr} k^{\alpha\pr}  
k^{\beta\pr}\nonumber
\end{eqnarray}
which is zero.

After rearranging the terms appearing in Eq.(\ref{a3}), and by the
application of Eqs.(\ref{pperpint}) and (\ref{primeperpint}) we
arrive at the expression
\begin{eqnarray}
\mbox{R}^{(E)}_{\mu \nu}\stackrel{\circ}{=}
4i\eta_{-}(p_0)\Bigg[\!\!\!\!\!\!\!&&\varepsilon_{\mu \nu
\alpha \beta} 
p^{\alpha} k^{\beta}(1 + \tan(eQ_f{\cal B}s)\tan(eQ_f{\cal B}s'))
+ \varepsilon_{\mu \nu \alpha
\beta\pr} k^{\alpha} k^{\beta\pr} \nonumber \\
&\times& \tan(eQ_f{\cal B}s) \tan(eQ_f{\cal B}s') \frac{\tan(eQ_f{\cal B}s)-\tan(eQ_f{\cal B}s')}{\tan(eQ_f{\cal B}s) + \tan(eQ_f{\cal B}s')}\Bigg].\nonumber\\
\label{evenpart}
\end{eqnarray}
Because of the presence of terms like $\varepsilon_{\mu \nu
\alpha \beta} 
 k^{\beta}$ and $ \varepsilon_{\mu \nu \alpha
\beta\pr} k^{\alpha} $ as  we contract $\mbox{R}^{(e)}_{\mu \nu}$  by
$k^\nu$, it vanishes. In view of this result it is tempting
to conclude that, axion coupling to
photon, in a magnetized medium, cannot be even in powers of external
magnetic field. 
{\it With all possibility this would be forbidden by the
discrete symmetries i.e CPT.}
 However, apart from the discrete symmetry arguments, 
there is an alternative. recalling the fact that each power of $eQ_f{\cal B}$
actually denotes insertion of very soft photon (i.e all components of 
the four vector, $k^{\lambda} \to 0$ ). 
In this case, the coupling of the dynamic photon along with 
the axion to even number of soft photon insertions, makes the power of
the electromagnetic vertex odd. Since axion has spin zero, in order 
to match the total spin of the system, the 
sum over photon spins should add up to zero, which is impossible with
odd number of photons. Hence this term should vanish in principle. 
Which is the result one arrives at after explicit computation.
%
%
%
%

\subsection{The Result for Odd Orders In Field Strength}

%
%
As has already been noted that, to odd orders in field strength one has
$R^{(O)}_{\mu\nu}$ and it can be expressed as a sum over two terms, one
proportional to $m^2$ and the other is $R^{(o)}_{\mu\nu}$, given by,
\begin{eqnarray}
\mbox{R}^{(o)}_{\mu\nu}&=& 4i\eta_{+}(p)\left[ m^2\varepsilon_{\mu \nu
1 2}(\tan(eQ_f{\cal B}s) + \tan(eQ_f{\cal B}s'))
+
\left\{(g_{\mu \alpha\para} p^{\widetilde{\alpha\para}}
p'_{\nu\para} - g_{\mu \nu} p'_{\alpha\para}
p^{\widetilde{\alpha\para}} +g_{\nu \alpha\para} p^{\widetilde{\alpha\para}}
p'_{\mu\para} )\right.\right.\nonumber\\
&+&\left.\left.(g_{\mu \alpha\para} p^{\widetilde{\alpha\para}}
p'_{\nu\pr} + g_{\nu \alpha\para} p^{\widetilde{\alpha\para}}
p'_{\mu\pr}) \sec^2(eQ_f{\cal B}s')\right\} \tan(e Q_f{\cal
B}s)\right.\nonumber\\
&+& \left.\left\{(g_{\mu \alpha\para} p'^{\widetilde{\alpha\para}}
p_{\nu\para} - g_{\mu \nu} p_{\alpha\para}
p'^{\widetilde{\alpha\para}} +g_{\nu \alpha\para} 
p'^{\widetilde{\alpha\para}}
p_{\mu\para} )
+
(g_{\mu \alpha\para} p'^{\widetilde{\alpha\para}}
p_{\nu\pr} + g_{\nu \alpha\para} p'^{\widetilde{\alpha\para}}
p_{\mu\pr}) \sec^2(eQ_f{\cal B}s)\right\}
\right.\nonumber \\ 
 &\times& \left.
 \tan(e Q_f{\cal
B}s') \right]. 
\label{rodd-1}
\end{eqnarray}
In this subsection, we would outline the proof that Eqn. [\ref{rodd-1}] on 
contraction with $k^{\mu}$ vanishes, for all $\nu$. The proof follows 
in two steps, first one demonstrates that for $\nu=1 \mbox{~or~}2$ it 
goes to zero. Followed by that one shows that the same vanishes
 for $\nu=0, \mbox{~or~}3$ as well. Denoting $\nu = 1$ or $\nu =2 $
collectively as , $\nu_{\perp}$, we show in appendix B,
\begin{eqnarray}
k^{\mu} \mbox{R}^{(o)}_{\mu \nu_{\perp}}(k)=0.
\end{eqnarray}
\noindent
Next  we  claim that, for $\nu= 0 ~or~3$, denoted collectively as 
$\nu_{\para}$
( i.e  the longitudinal components); 
$k^{\mu}R^{(o)}_{\mu\nu_{\para}}$ would vanish.
%
%
One can verify, with a bit of algebra,  that
the same can be written as a sum over two pieces.
That is,\\ 
$$
k^{\mu}R^{(o)}_{\mu\nu_{\para}}= 
k^{\mu}R^{(o.1)}_{\mu\nu_{\para}}+k^{\mu}R^{(o.2)}_{\mu\nu_{\para}} $$
Apart from the uninteresting
overall constants and integrations over the variables
p,p', s and s' the pieces $k^{\mu}R^{(o.1)}_{\mu\nu_{\para}}$ are 
found to be proportional to,
\begin{equation}
k^{\mu}R^{(o.1)}_{\mu\nu_{\para}}\!\!\! =
 p_{\widetilde{\nu_{\para}}}\!\! \left[
\left( k^2_{\para} +2 (k \cdot p)_{\para}\!\!\right)\left( \tan(eQ_f{\cal B}s) +\tan(eQ_f{\cal B}s')
\right)
- k^2_{\perp} (\tan(eQ_f{\cal B}s)- \tan(eQ_f{\cal B}s'))\right].
\label{pal1}
\end{equation}
The fact that this piece vanishes exactly on general symmetry grounds
of the integrals was first noted in \cite{faraday} and also in 
\cite{neutrinoII}, so we skip the details till appendix C. The required  
elaboration is provided there.\\

\noindent
 The left over pieces, not 
considered yet, is $ k^{\mu}R^{(o.2)}_{\mu \nu_{\para}}$. That is given by,
\begin{eqnarray}
k^{\mu}R^{(o.2)}_{\mu 0}= - 4i\eta_{+}(p)k^3\left[\left(p^2_{\para}-m^2 \right)
\left( \tan(eQ_f{\cal B}s)\!\!\! \right.\right.\!\ &+&\left.\left.\!\! \tan(eQ_f{\cal B}s')\right) - 
k^2_{\perp} \tan^2(eQ_f{\cal B}s') \right.\nonumber \\ &\times& \left.
\frac{\sec^2eQ_f{\cal B}s}{\tan(eQ_f{\cal B}s)
+ \tan(eQ_f{\cal B}s')}\right] 
\label{vertex000}
\end{eqnarray}
Similarly,
\begin{eqnarray}
 k^{\mu}R^{(o.2)}_{\mu 3}  = 4i\eta_{+}(p)k_0\left[\left(p^2_{\para}-m^2 
\right)
\left( \tan(eQ_f{\cal B}s)\!\!\! \right.\right. &+& \left. \left.
 \tan(eQ_f{\cal B}s')\right) - k^2_{\perp} \tan^2(eQ_f{\cal B}s')
\right.\nonumber \\ &\times& \left.
\frac{\sec^2eQ_f{\cal B}s}{\tan(eQ_f{\cal B}s)+ \tan(eQ_f{\cal B}s')}\right] 
\label{vertex3aaa}
\end{eqnarray}
In appendix D we have shown that Eqns [\ref{vertex000}] and [\ref{vertex3aaa}]
vanishes.Thus the proof that,
$k^{\mu}R^{(o)}_{\mu\nu}(p,p',s,s')=0$ is complete.
\section{To The Effective Interaction.}
\noindent
So using the results of the previous section, the effective 
axion photon vertex can be written down. Noting that,
\begin{eqnarray} 
 k^{\mu}{\cal{R}}^{(O)}_{\mu\nu} 
=
- 8im^{2}\eta_{+}(p)
\left[  k^{\mu}\varepsilon_{\mu \nu 1 2}(\tan(eQ_f{\cal B}s) 
+ \tan(eQ_f{\cal B}s'))\right],
\end{eqnarray}
the vertex function $\Gamma_{\nu}(k)$ ( using Eqn. [\ref{compl}])
turns out to be,
\begin{eqnarray} 
\Gamma_{\nu}(k) = (g_{af}e Q_f)\left(8m^{2}
k^{\mu}\varepsilon_{\mu \nu 1 2} \right)
\int\!\!\frac{d^4p}{(2\pi)^4}\eta_{+}(p) \int^{\infty}_{-\infty}\!\!
\!\!\!\!\! ds
\int^{\infty}_{0}\!\!\!\!\!\!\! \!ds' \!\!&&\!\!\!\! e^{\Phi(p,s)+\Phi(p',s')}
 \nonumber \\
 &\times& \left[ \tan(eQ_f{\cal B}s) 
+ \tan(eQ_f{\cal B}s') \right]
\label{Thevertex}
\end{eqnarray}
Since the perpendicular components of the momentum integrals
are Gaussian, one can carry them out without any difficulty with the following
 result,
\begin{eqnarray}
\Gamma_{\nu}(k) = (g_{af}e Q_f)\left(8m^{2}
k^{\mu}\varepsilon_{\mu \nu 1 2} \right) &&\!\!\!\!\!\!\!\!\!\!
\int\!\!\frac{d^2p_{\para}}{(2\pi)^2}\eta_{+}(p) 
\int^{\infty}_{-\infty}\!\!\!\!\!\!\! ds
\int^{\infty}_{0}\!\!\!\!\!\!ds'  e^{is({p^2}_{\para}-m^2))
+is'({p'^2}_{\para}-m^2)) }
 \nonumber \\
 &\times& \frac{ -ieQ_f{\cal B}
\left[ \tan(eQ_f{\cal B}s) 
+ \tan(eQ_f{\cal B}s') \right]}{4 \pi \left[ \tan(eQ_f{\cal B}s) 
+ \tan(eQ_f{\cal B}s') \right] }
e^{-i\frac{ k^2_{\perp}}{eQ_f{\cal B}}\frac{ \tan(eQ_f{\cal B}s)\tan(eQ_f{\cal B}s')}
{  \left[ \tan(eQ_f{\cal B}s) 
+ \tan(eQ_f{\cal B}s') \right]}} \nonumber \\
\label{perpIntegrated-Thevertex}
\end{eqnarray}
Exact evaluation of [\ref{perpIntegrated-Thevertex}] is extremely difficult,
however it is possible to get some  analytical results under certain
approximations. In the long wavelength limit and for $\omega < m_f$ one can
get an analytical estimate of the same. However for energies above the
Pair Production Threshold (PPT), the one has to be very careful in evaluating 
the vertex functions. In the passing we note that, 
 in the limit of $m > \mu$ Eqn.
 [\ref{perpIntegrated-Thevertex}], can be evaluated analytically using
Poisson summation formula and  Laguerre polynomials. However in the limit of
vanishing $|k_{\perp}|$, the leading order result can easily be recovered from
eqn. [\ref{perpIntegrated-Thevertex}] by setting setting $|k_{\perp}| \to 0$ 
in the same.
The issue of  evaluation of the same for all values of $\omega$ and $\vec{k}$
 would be taken up in a separate publication. 
Since, in this work,  we would like to focus our attention to photon
energies below pair production threshold. In what follows, from now on, we
assume photon energy to be below PPT and would take the long wave length 
limit, i.e $|k_{\perp}| \to 0$, and try
to estimate the magnitude of the contribution of the vertex function.
\begin{eqnarray}
\lim_{|k_{\perp}| \to 0} \!\!\!\Gamma_{\nu}(k) &=& (g_{af}e Q_f)\left(8m^{2}
k^{\mu}\varepsilon_{\mu \nu 1 2} \right)\frac{\! -ieQ_f{\cal B}}{4 \pi }\!
\int\!\!\frac{d^2p_{\para}}{(2\pi)^2}\eta_{+}(p)\!\!
\int^{\infty}_{-\infty}\!\!\!\!\!\!\!\! ds
\int^{\infty}_{0}\!\!\!\!\!\! ds'  e^{is(p^2_{\para}-m^2))
+is'(p'^2_{\para}-m^2)) }
 \nonumber \\ 
&=& -(ig_{af}(eQ_f)^2)\left(4 m^{2}
k^{\mu}\varepsilon_{\mu \nu 1 2} \right){\cal B} 
\int\!\!\frac{d^2p_{\para}}{(2\pi)^2}\eta_{+}(p) 
\delta (p^2_{\para}-m^2))\frac{-i}{(p'^2_{\para}-m^2)}
\label{perpIntegrated-Thevertex1}
\end{eqnarray}
The last step in Eqn.[\ref{perpIntegrated-Thevertex1}]follows from the 
definition of the delta function preceded by  subsequent
integration over $s^{\prime}$.
\begin{eqnarray}
\int^{\infty}_{-\infty}\!\!\!
e^{is(p^2_{\para}-m^2)} ds &=& 2\pi \delta (p^2_{\para}-m^2)) 
\nonumber  \\ 
\int^{\infty}_{0} \!\!\!
e^{is'(p'^2_{\para}-m^2)} ds' &=&  \frac {-i}{(p'^2_{\para}-m^2)}
\label{eval1}
\end{eqnarray}
In the rest frame of the medium $p.u= p_0$, in this frame, the thermal factor,
$\eta_{+}(p.u)$ turns out to be, $\eta_{+}(p_0)= n_F(|p_0|,\mu) 
+ n_F(|p_0|,-\mu) $ \cite{faraday}, (Here $n_F(x)$, the thermal distribution 
function, is defined as $n_F(x,\mu)= \frac{1}{e^{\beta(x -\mu)}+1}$, as 
is the case for Fermi distribution fn.). using the same and the delta function
constraint, one can simplify Eqn. [\ref{perpIntegrated-Thevertex1}] further,
to arrive at, 
\begin{eqnarray}
\Gamma_{\nu}(k)&=&\!\!-(g_{af}(eQ_f)^2)\left(4 m^{2}
k^{\mu}\varepsilon_{\mu \nu 1 2} \right)\! {\cal B} \!\!\!
\int\!\!\frac{d^2p_{\para}}{(2\pi)^2}\Big[ n_F(|p_0|,\mu) 
+ n_F(|p_0|,-\mu) \Big] \nonumber \\
&& \mbox{\hskip 5cm}\!\!\times\delta (p^2_{\para}-m^2)\frac{1}{(k^2_{\para}+ 2 (p\cdot k)_{\para})}
\nonumber \\ 
&=& -16 (g_{af}(eQ_f)^2)
\left( \frac{k^{\mu}\tilde{\mbox{F}}_{\mu \nu}}
{16\pi^2} \right) \Lambda(k^2_{\para},
k\cdot u, \beta, \mu)
\label{perpIntegrated-Thevertex2}
\end{eqnarray}
Where $\Lambda(k^2_{\para},k_{\para}\cdot u, \beta, \mu  ) $ is,
\begin{eqnarray}
 \Lambda(k^2_{\para},k \cdot u, \beta, \mu  )\!\! &=& \!\!\!\!
\int\!\!\mbox{d}^2p_{\para} \Bigg[ n_F(|p_0|,\mu) 
+ n_F(|p_0|,-\mu) \Bigg] \!\!
\left(\frac{ m^2\delta (p^2_{\para}-m^2)}{(k^2_{\para}+ 2 (p\cdot k)_{\para})}
\right)
\label{perpIntegrated-Thevertex3}
\end{eqnarray}
%
%
Therefore the effective 
Lagrangian for axion photon interaction would be given by.
\begin{eqnarray}
{\cal L}^{{\cal B},\mu,\beta}_{\gamma a}&=& A^{\nu}\Gamma_{\nu}(k)= - 16 \left(g_{af}(eQ_f)^2 \right)
\frac{\mbox{F}_{\mu\nu}\tilde{\mbox{F}}^{\mu\nu}}{16\pi^2}
\Lambda ( k^2_{\para},k \cdot u, \beta, \mu ).
\label{efflag}
\end{eqnarray}  
Equation [\ref{efflag}] is the expression for 
axion photon coupling in a magnetized medium, in the limit, 
$|k_{\perp}| \to 0$. It is important to note that,
 this equation depends on  many physical parameters,
e.g., the temperature of the medium ( $\beta =1/T $), number 
density of the 
fermions (which in turn is related to $\mu$), mass of the particles 
in the loop (m), energy and longitudinal momentum of the photon
( i.e. $k_{||}$) and of course the symmetry 
breaking scale ( which
is included in the coupling constant, $g_{af}$). \\

\noindent
Since the basic purpose 
of this exercise is to find out the effect of the medium on the mixing
of axions to photons, we would have to choose some energy scale for
the parameters entering into the statistical factors. Since most of the
studies in literature deals with optical photons, we would assume
the following scale, i.e  $m_f \gg k^2 $ and finally
$\omega > k_3 $. We would like to point out that, $\omega \equiv k_0 $ 
and $k_3$ would henceforth be denoted as $k$. 
Armed with these we can
now turn to evaluate, Eqn. [\ref{perpIntegrated-Thevertex3}].In the limit as mentioned above.
\begin{eqnarray}
 \Lambda(k^2_{\para},k \cdot u, \beta, \mu  ) 
&=& 
- \int^{\infty}_{-\infty}\!\!dp \left[ \frac{ n_F(E_p,\mu) +n_F(E_p,-\mu)}{2E_p} \right] 
\left[m^2 \cdot \frac{2 k^2_{\para}}{4\omega^2 E^2_p}\right].  
\label{perpIntegrated-Thevertex4}
\end{eqnarray}
 We  would like to point out that, 
while coming to the last step in equation [\ref{perpIntegrated-Thevertex4}],
we first had performed the the $p_0$, integral, then had expanded the 
resulting denominator, in powers of  $\Big[\frac{2p.k -k^2_{\para}}
{2\omega E_p}\Big]$,  while retaining the leading order term in powers of 
$\Big[\frac{k^2_{\para}}{\omega E_p}\Big]$ . In the expressions above
$E_p = \sqrt{(p^2+m^2)}$.

\begin{eqnarray}
 \Lambda(k^2_{\para},k \cdot u, \beta, \mu  )
 = - \left(\frac{k_{\para}}{\omega}\right)^2 \int^{\infty}_{-\infty}\!\! dp 
\Big[ \frac{ n_F(E_p,\mu)
+n_F(E_p,-\mu)}{2E_p}\Big] \frac{m^2}{2 E^2_p}.  
\label{perpIntegrated-Thevertex5}
\end{eqnarray}
%
From now on we denote $
\Lambda  =  -\int^{\infty}_{-\infty}\!\! dp  \left[\frac{n_F(E_p,\mu)
+n_F(E_p,-\mu)}{2E_p}\right] \frac{m^2}{2 E^2_p}.$
If we  neglect pieces 
proportional to $k^2/\omega^2$, its easy to see that  
$ \Lambda(k^2_{\para},k \cdot u, \beta, \mu  ) \equiv \Lambda =-\int^{\infty}_{0}dp
 \left[\frac{n_F(E_p,\mu)
+n_F(E_p,-\mu)}{2E_p}\right] \frac{m^2}{2 E^2_p} $. Other wise (for small $k \& k_{\para}$),
 $\Lambda(k^2_{\para},k \cdot u, \beta, \mu  ) = \left(\frac{k_{\para}}{\omega}\right)^2
\!\!\! \Lambda$.

\subsection{Limiting Case: $m \gg  \mu$. }
%
Analytical evaluation of eqn. [\ref{perpIntegrated-Thevertex5}] for arbitrary 
values of chemical potential $\mu$ and temperature is an extremely difficult 
task. However as has already been mentioned, under some approximations it 
may be possible to get an analytical
form for the same. One such approximation is $m \gg \mu$. This is reasonable
in most of the astrophysical and all of cosmological situations, 
barring the core of the Supernova
or active galactic nuclei where this approximation may not hold. 
In this approximation,( i.e $m \gg \mu$)  thermal Fermi-Dirac factors can 
 be expanded in a series to give,
\begin{eqnarray}
 n_F(E_p,\mu) + n_F(E_p,-\mu)= 2 \sum^{\infty}_{n = 0} (-1)^n 
\cosh\left[(n+1)\beta\mu\right]\, e^{-(n+1)\beta E_p}.
\label{Fddist}
\end{eqnarray}  
In order to evaluate, Eqn. [\ref{perpIntegrated-Thevertex5}], we would
 substitute Eqn. [\ref{Fddist}] in the same eqn., to arrive at,
\begin{eqnarray}
 \Lambda
& = & - \sum^{\infty}_{n = 0} (-1)^n\cosh\left[(n+1)\beta\mu\right]
\int\!\!  dp \frac{e^{-(n+1)\beta E_p}}{E_p} 
\frac{m^2}{ E^2_p}. 
\label{perpIntegrated-Thevertex6}
\end{eqnarray}
The evaluation of Eqn.[\ref{perpIntegrated-Thevertex6}] is a bit difficult,
however, expression appears a bit less formidable, if one  identifies,
 $\sqrt{s}$ with \footnote{It should noted that this s is 
different from the proper time parameter s introduced in the fermionic 
Schwinger propagators.}
$E_p$ and uses the following integral transform,
\begin{eqnarray}
s^{-\frac{1}{2}}e^{-\alpha \sqrt{s}}= \pi^{-\frac{1}{2}}
\int^{\infty}_{0} e^{-us - \frac{\alpha^2}{4u}}\frac{du}{\sqrt{u}}
\label{transf1}
\end{eqnarray}
 to convert Eqn.[\ref{perpIntegrated-Thevertex6}] close to a Gaussian. 
Upon using this transform, the p dependent part of the integrand
goes over to,
\begin{equation}
2 \int dp \frac{e^{-p^2u}}{p^2+ m^2} = \frac{\pi}{m}\left[
1- \Phi(m \sqrt{u}) \right]e^{um^2}.
\label{errf}
\end{equation}
Here the function $\Phi(m \sqrt{u})$ stands for error function. Fortunately
there exists another integral representation for the right hand side in
Eqn. [\ref{errf}], given by,
\begin{equation}
2 \frac{\sqrt{\pi}}{m} \int dt e^{-t^2- 2tm\sqrt{u}} = \frac{\pi}{m}\left[
1- \Phi(m \sqrt{u}) \right]e^{um^2}.
\label{errf1}
\end{equation}
Use of this expression makes the expressions manageable. Using eqn.[\ref{errf1}], and expanding
the left hand side in powers of
$u$ and then performing the t and u integration one arrives at,
\begin{eqnarray}
\Lambda &=& -\frac{m^2}{2 m}\sum^{\infty}_{n=0} (-1)^n cosh[(n+1)\beta\mu]
 \sum^{\infty}_{l=0} \frac{(-1)^l(2m)^l}{l!}
\int^{\infty}_{0} \frac{du}{\sqrt{u}} (\sqrt{u})^{l} e^{-m^2u - 
\frac{(n+1)\beta}{4u}}
\nonumber \\
&& \times  \int^{\infty}_{0} e^{-t^2} t^{l} dt.
\label{L0}
\end{eqnarray}
In Eqn.[\ref{L0}], u integration would result in giving modified Bessel 
function and following that,
the t integration, can be performed using the formula provided below,
\begin{eqnarray}
\int^{\infty}_{0} x^{\nu -1} e^{-\mu x^p} = \frac{1}{p}\mu^{-\nu/p}
\Gamma\left(\frac{\nu}{p}\right).
\end{eqnarray}
to arrive at the following expression for $\Lambda$.
\begin{eqnarray}
\Lambda = -\frac{1}{2}\sum^{\infty}_{n=0} (-1)^n \cosh[(n+1)\beta\mu]
 \sum^{\infty}_{l=0} \frac{(-1)^l(2)^{l+1}}{l!}
\Gamma(\frac{l+1}{2})
\nonumber \\ ( \frac{(n+1)\beta m}{2})^{\frac{l+1}{2}}
K_{\frac{(l+1)}{2}}[(n+1)m\beta].
\label{intdone}
\end{eqnarray}

One can use the expression [\ref{intdone}] in  equation [\ref{efflag}] and 
estimate the contribution of magnetized medium to photon axion mixing. 
For simple cases, e.g. in high temperature (more relevant for the
cosmological and astrophysical scenario) or low temperature (intergalactic 
regions)  it should be possible to use different expansions for the modified
Bessel functions and estimate the size of the corrections to arrive at 
the analytical results.\\

%
\subsection{Limit: $m \le \mu \le \infty$}
%
In certain physical situations matter density can become extremely high,
i.e the highly degenerate Fermi system. Examples are nascent neutron
stars or the core of the core collapse supernova where matter density
can be $10^{15}\frac{gm}{cc} $ or more and temperature can vary 
between several tens of MeV to few KeV. Keeping this physical situation 
in mind we would try to estimate the contribution of $\Lambda$ for 
various values of chemical potential. Noting that the $a\gamma$ effective 
Lagrangian can be written in the form 
%
\begin{eqnarray}
{\cal L}^{{\cal B},\mu,\beta}_{\gamma a} &=& 
\frac{a \mbox{F}\tilde{\mbox{F}}}
{64\pi^2}
g_{af}(eQ_f)^2 \left(\frac{k_{\para}}{\omega} \right)^2
32\cdot \left[ \int^{\infty}_{0} dp \frac{1}{\left(p^2+1\right)^{\frac{3}{2}}}  
\left\{ \frac{1}
{e^{\beta m \left( p^2 + 1 \right)^{\frac{1}{2}} -\beta\mu} +1} 
 + (\mu \leftrightarrow -\mu )  \right\}\right] \nonumber \\
&=& \frac{a \mbox{F}\tilde{\mbox{F}}}
{32\pi^2}
g_{af}(eQ_f)^2
\left(\frac{k_{\para}}{\omega} \right)^2
16\cdot \widetilde{\Lambda} 
\label{degleff}
\end{eqnarray}
%
The integral inside square bracket in Eqn. [\ref{degleff} ] is being
denoted by $\widetilde{\Lambda}$ where $\widetilde{\Lambda}= - 2 \Lambda $. 
It is worth noting that inside the 
integral, the function inside the braces for $m \ll \mu$ and 
$\rm{limit}{\rm{T} \to 0}$ can be approximated by $\Theta( \frac{\mu}{m} -p )$. 
In this limit we can write, 
\begin{eqnarray}
\rm{Lt}_{T \to 0 }\widetilde{\Lambda} \simeq \int^{\mu \over m}_{0} 
\frac{dp}{\left( p^2+ 1\right)^{3/2}}= 
\rm{sin}\left[ \rm{tan}^{-1}\left(\frac{\mu}{m} \right) \right] 
=\frac{\frac{\mu}{m}}{\sqrt{\left(1+ \left[\frac{\mu}{m}\right]^2 \right)}}
\label{analytical}
\end{eqnarray}
In the limit $\mu \gg m$, the right hand side of Eqn. (\ref{analytical})
$\sim$ 1.

\begin{figure}
  \begin{center}
    \input{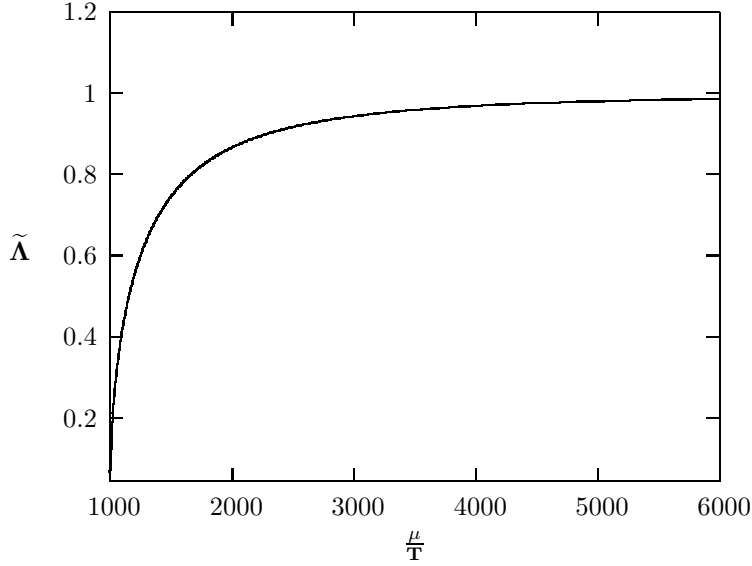}
   \end{center}
\caption[]{$\widetilde{\Lambda}$ vs. $\frac{\mu}{T}$ for  relativistic degenerate medium. The parameters are as follows: $m \le \mu \le 6$ when the temperature
is held fixed at, $T=10^{-3}$ in units of fermion mass. }
\label{Fig:f2}
\end{figure}

Eqn. [\ref{degleff}] has also been
estimated numerically, for $m \le \mu \le 6$ when the temperature
is held fixed at, $T=10^{-3}$ in units of fermion mass. The behavior
of the function $\widetilde{\Lambda}$ in this range can be seen in 
[Fig.\ref{Fig:f2} ] which saturates at $0.98$ providing good agreement
between analytical and numerical result.\\
%

\noindent
The important part of this analysis is, for sufficiently large values of
matter density and low temperature the value of $\Lambda$ is seen to
saturate at a value not exceeding one. Though for smaller values of matter
density it ranges between half to unity. 
%
\section{Discussion and Conclusion}
As has already been mentioned at the beginning, the purpose of this note
is to find out the modifications in the axion photon vertex in the presence
of an external magnetic field and matter. Field induced part of the
photon axion vertex has been worked out earlier in \cite{raff-m}, however
the effect of magnetized matter wasn't taken into account. In this note
we have considered the effect of both external magnetic field and medium
on the vertex. More over we have also written down the explicit formula for 
the vertex involving the PQ charges of the fermions. In the light of these 
estimates,  it is possible to write down the axion photon mixing Lagrangian,
for low frequency photons in an external magnetic field, in the following way:
\begin{eqnarray}  
{\cal L}^{Total}_{a \gamma} = {\cal L}^{vac}_{a \gamma}+
{\cal L}^{{\cal B}}_{a \gamma}+ {\cal L}^{{\cal B},\mu,\beta}_{a \gamma}.
\end{eqnarray}
Where ${\cal L}^{vac}_{a \gamma}$ is the usual axion photon mixing term 
in vacuum, resulting form the electromagnetic anomaly, and is given by
\cite{kim},
\begin{eqnarray}
\mbox{\hskip 3.25 cm}
{\cal L}^{vac}_{a \gamma}&=& - g_{a \gamma \gamma} \frac{e^2}{32\pi^2}
a \mbox{F}{\tilde{\mbox{F}}}, \mbox{~~where~} g_{a \gamma \gamma}= 
\frac{A^{c}_{PQ}}{f_a} \left[\frac{A^{em}_{PQ}}{A^{c}_{PQ}}- 
\frac{2(4+z)}{3(1+z)}\right]
\!\!\!\!\!\!\! \nonumber \\ 
%
{\cal L}^{\cal B}_{a\gamma}& =& \frac{-1}{32 \pi^2 } 
\left[ 4 + \frac{4}{3} \left( \frac{k_{\para}}{m} \right)^2 \right]
\!\!\!\sum_f
g_{ af} (eQ_f)^2 
a \mbox{F}\tilde{\mbox{F}}. \nonumber \\ 
{\cal L}^{{\cal B},\mu,\beta}_{\gamma a}&=&\frac{16}{32\pi^2} \cdot
\left(\frac{k_{\para}}{\omega}\right)^2 ({\widetilde\Lambda}) \sum_f 
g_{af}(eQ_f)^2
a\mbox{F}\tilde{\mbox{F}}.
\label{axpefl}
\end{eqnarray}   
%
To remind ourselves,
in the equations above, $z=\frac{m_u}{m_d}$, where $m_u$ and $m_d$ stands for the masses of the light quarks. 
The anomaly factors are given by the following
relations, 
$A^{em}_{PQ} = \mbox{Tr}(Q^2_f)X_{f}$ and $\delta_{ab}A^{em}_{c}= 
\mbox{Tr}\lambda_a\lambda_b X_{f}$ (where the trace is over the fermion 
species.). 
Hence, in the limit of $|k_{\perp}| \to 0 $ and $\omega << m_f $, using 
eqn. [\ref{axpefl}], one can write the total axion photon effective Lagrangian:
 
\begin{eqnarray}
{\cal L}^{Total}_{a \gamma}&=& - \left[g_{a \gamma \gamma} 
+ \left( 4 + \frac{4}{3}\left(\frac{k_{\para}}{m} \right)^2 \right)
\sum_f g_{af}(Q_f)^2 
-16 \left(\frac{k_{\para} }{\omega}\right)^2 ({\widetilde\Lambda}) \sum_f 
g_{af}(Q_f)^2 \right] \frac{e^2}{32\pi^2}
a\mbox{F}\tilde{\mbox{F}}.
\label{axpefl.b}
\end{eqnarray} 
In order to describe the axion photon interaction, for static magnetic field
and real photon, the effective Lagrangian employed in the literature is 
usually given by the following relation,
\begin{eqnarray}
\frac{1}{M}a \vec{\mbox{E}}.\vec{\cal{B}}^{ext}.
\label{usualeffectivelag}
\end{eqnarray}
Where the parameter $M$, defines an energy scale in terms of inverse of the 
 axion photon photon coupling constant in vacuum, i.e 
$ M \equiv \frac{32\pi^2}{e^2 g_{a \gamma \gamma}}$. In principle this factor is a 
model dependent quantity, and depending on the particular choice of PQ 
charges, $g_{a\gamma \gamma}$ can vary between zero to hundred. This 
is one of the model dependent uncertainty. On the other hand,
in some experiments, attempts have been made to give a bound on
$M$ thus $g_{a\gamma \gamma}$ through the detection of solar axions
in a magnetic cavity haloscope. The astrophysical bounds on $M$ come from
the constraint that the stars don't lose energy through axion photon 
conversion at a rate faster than the generation of energy. As a result
of these investigations, the present day bound on the energy scale
 is believed to be, $M \le 10^{12} GeV$.  On the other hand
in this note, we have been able to point out that apart from the
model dependent uncertainties, there is also medium dependent uncertainties
that can introduce some variation on
 the bounds on $M$.
The amount of this  variation would depend on the kind of environment
one is interested in. 
 The details of its implications for various models
and physical situations are extremely interesting by their own right
\cite{massimo,cameron,semertzidis,soa,long,stod,wilc,miani,gas,par};
however these points are beyond the scope of this work.
%
%
%
%
\section*{Acknowledgment}
\noindent
Author wishes to thank Prof. P.~Aurenche and Prof. A.~Roychowdhury
and Prof. P.~B.~Pal for their interest and encouragement. It is also a pleasure to thank 
the director Haldia Institute Of Technology, Prof.~T.~S. Banerjee, 
Prof. A.~D.~Mukhopadhyay and  Secratary ICARE, Mr. L.~Seth and his 
colleagues for creating  the right kind of  ambience.
\renewcommand{\thesection}{Appendix}
\renewcommand{\theequation}{\Alph{section}.\arabic{equation}}
\section*{appendix} 
\subsection*{Appendix A :  The Equations.}

Contribution from magnetized material part to $VA$ response function
comes through the pure thermal propagators (in the external magnetic field).
 Eqn.[\ref{pi-ini}] reads:
\begin{eqnarray}
i\Pi^{A_{\beta B}}_{\mu\nu}(k)=(-i e^2)(-1) \int {{d^4 p}\over {(2\pi)^4}}\mbox{Tr}\left
[ \gamma_\mu \gamma_5 S^\eta_B(p) \gamma_\nu iS^V_B(p') + \gamma_\mu
\gamma_5 iS^V_B(p) \gamma_\nu S^\eta_B(p')\right]
\label{pi-iniapp}
\end{eqnarray}
Eqn.[\ref{pi-iniapp}] has two pieces in it. In what follows we would try to 
demonstrate that except the thermal weight factors (i.e the $\eta_F(p)$s)
the traces of both the terms come out to be the same.\\

\noindent
To begin with, we would concentrate on the first piece.
In terms of the functions $G(p,s)$ and $G(p',s')$ the same can be 
written as,
\begin{eqnarray}
\mbox{Tr}\left[\gamma_\mu \gamma_5 S^\eta_B(p) \gamma_\nu
iS^V_B(p')\right]
&=& -\mbox{Tr}\left[ \gamma_\mu \gamma_5 \int_{-\infty}^{\infty} ds\,
e^{\Phi(p,s)} G(p,s) \gamma_\nu \int_0^\infty ds'\,
e^{\Phi(p',s')} G(p',s')\right]\eta_F(p)\nonumber\\
&=& -\int_{-\infty}^\infty ds\, e^{\Phi(p,s)} \int_0^\infty ds'\,e^{\Phi(p',s')} \mbox{Tr}\left[\gamma_\mu
\gamma_5 G(p,s) \gamma_\nu G(p',s')\right]\eta_F(p)
\label{pi-part}
\end{eqnarray}
We would like to note the appearance of the integration variables,$p,p'$ and
 $s,s'$. \\
 
\noindent
The second term in the right hand side of eqn.(\ref{pi-iniapp})  needs some 
formal mathematical manipulations. Its better to do the same in two steps.
The first step would be {\it{shift the integration variable
 $p \to p-k $}}. And following this the next step would be to
{\it{change the integration variable $p \to -p $}}. As a result of this 
manipulation one arrives at,
\begin{eqnarray}
\mbox{Tr}\left[\gamma_\mu \gamma_5 iS^V_B(p) \gamma_\nu
S^\eta_B(p')\right]&=&\mbox{Tr}\left[\gamma_\mu \gamma_5 iS^V_B(-p') \gamma_\nu S^\eta_B(-p)\right]\nonumber\\
&=& -\mbox{Tr}\left[ \gamma_\mu \gamma_5 \int_0^{\infty} ds'\,
e^{\Phi(-p',s')} G(-p',s') \gamma_\nu \int_{-\infty}^\infty ds\,
e^{\Phi(-p,s)} G(-p,s)\right]\eta_F(-p)\nonumber\\
&=& -\int_{-\infty}^\infty ds\, e^{\Phi(p,s)} \int_0^\infty ds'e^{\Phi(p',s')} \mbox{Tr}\left[\gamma_\mu
\gamma_5 G(-p',s') \gamma_\nu G(-p,s)\right]\eta_F(-p)
\label{pi-part2}
\end{eqnarray}
 The complete expression is given by,
\begin{eqnarray}
i\Pi^{A_{\beta B}}_{\mu\nu}(k)&=&-(-i e^2)(-1) \int {{d^4 p}\over {(2\pi)^4}}
\int_{-\infty}^\infty ds\, e^{\Phi(p,s)} \int_0^\infty
ds'\,e^{\Phi(p',s')}\nonumber\\
& &\times \left[ \mbox{Tr}\left[\gamma_\mu\gamma_5 G(p,s)
\gamma_\nu G(p',s')\right]\eta_F(p) + \mbox{Tr}\left[\gamma_\mu
\gamma_5 G(-p',s') \gamma_\nu G(-p,s)\right]\eta_F(-p)
 \right]
\label{complapp1}
\\
 &=& -(-i e^2)(-1) \int {{d^4 p}\over {(2\pi)^4}}
\int_{-\infty}^\infty ds\, e^{\Phi(p,s)} \int_0^\infty
ds'\,e^{\Phi(p',s')}\quad \mbox{R}_{\mu\nu}(p,p',s,s')
\label{complapp}
\end{eqnarray}
where $\mbox{R}_{\mu\nu}(p,p',s,s')$ contains the trace part. It is worth 
noting that the two pieces appearing in the second line
 of Eqn. [\ref{complapp}] are related
to each other by the  following transformation $\left(p \to -p' \right)
$ and $\left( s \to s' \right)$. So once we compute the trace of one of them,
the other can easily be obtained by interchanging the variables,$p \to -p'$
 and $s \to s'$.

\noindent
Hence, in the following we would evaluate just one of them. If we introduce the 
following short hand notations,
\begin{eqnarray}
\tan(eQ_f{\cal B}s) = T_s    & \mbox{~~~~~and~~~~~} & \sec^2(eQ_f{\cal B}s) = S_s \nonumber \\
\tan(eQ_f{\cal B}s') = T_{s'}& \mbox{~~~~~and~~~~~} & \sec^2(eQ_f{\cal B}s') = S_{s'}\nonumber \\
\Gamma_{\mu 5}=\gamma_{\mu}\gamma_{5}.
\label{short}
\end{eqnarray}
The expression for, $G(p,s)$ can further be written as, 
\begin{eqnarray}
G(p,s) &=& (1+i \sigma_z \mbox{tan}(eQ_f{\cal B}s))(\rlap/p_\parallel+m)-\mbox{sec}^2(eQ_f{\cal B}s)\rlap/p_\perp \\ \nonumber
&=&
(1+i \sigma_z T_s)(\rlap/p_\parallel+m)- S_s \rlap/p_\perp
\label{G}
\end{eqnarray}
After doing a bit of  algebra, the Dirac trace part of the first term
of Eqn. [\ref{complapp1}], odd in powers of $eQ_f{\cal B}$ turns 
out to be (written in terms of short-hand notations introduced in 
Eqn.[\ref{short}]),
\begin{eqnarray}
i\hat{\Pi}^{(A_{\beta B})-1.}_{\mu\nu} &=&- (-i)^2(g_{af}eQ_f)(-1)\int \frac{d^4p}{(2\pi)^4} 
\int^{\infty}_{-\infty} e^{\Phi(p,s)}ds \int^{\infty}_{0} ds' 
e^{\Phi(p',s')}
\\ \nonumber
&\times &
i\left[ \Gamma_{\mu 5}\left(
\sigma_z \rlap/p_{\para}\gamma_{\nu}\rlap/p'_{\para} T_s+
\rlap/p_{\para}\gamma_{\nu} \sigma_z\rlap/p'_{\para} T_s'
\right) +
m^2 \left( \Gamma_{\mu 5}\sigma_z\gamma_{\nu}T_s +
\Gamma_{\mu 5}\gamma_{\nu} \sigma_z T_s \right) \right. \nonumber \\
\left. 
 \right. &+& \left. \Gamma_{\mu 5} \sigma_z\rlap/p_{\para} \gamma_{\nu} \sigma_z\rlap/p'_{\perp}
T_sS_s' +
\Gamma_{\mu 5}\rlap/p_{\perp}\gamma_{\nu} \sigma_z\rlap/p'_{\para}
T_{s'}S_s
\right]\eta_F(p)
\label{trace1}
\end{eqnarray}
In order to simplify things a bit, we note that, using the defn of $\sigma_z$,
(i.e $\sigma_z= - \gamma_0\gamma_3\gamma_5 $)one can write,
\begin{eqnarray}
\sigma_z \rlap/p_{\para}= i\gamma_1\gamma_2 \rlap/p_{\para} &= &
\gamma_0\gamma_3 \rlap/p_{\para}\gamma_{5}=
\gamma_0\gamma_3\left(\gamma_0p^{0} +\gamma_{3}p^3 \right) \gamma_{5}
=-\left(\gamma_0p^{3}+ \gamma_3 p^{0} \right) \gamma_{5} 
\nonumber \\
&=& -\gamma_{\alpha_{\para}}p^{\widetilde{\alpha_{\para}}},
\mbox{\hskip 2.5cm ($\alpha_{\para}$ could be, either $0$ or $3$)}.
\label{tilde}
\end{eqnarray}
We would also like to point out that, depending on the value of $\alpha_{\para}$, $\widetilde{\alpha_{\para}}$ would
take its complimentary value, i.e to say, if 
$\alpha_{\para}=0$ then $\widetilde{\alpha_{\para}}=3$ etc. 
We also note in the passing that in this notation,
\begin{eqnarray}
\sigma_z\rlap/p'_{\para} &=& - \gamma_{\alpha_{\para}}\gamma_5 p^{'
\widetilde{\alpha_{\para}}}.
\nonumber \\
\sigma_z\rlap/p_{\para} &=& - \gamma_{\alpha_{\para}}\gamma_5 p^{
\widetilde{\alpha_{\para}}} \mbox{\hskip 3cm and ~finally,}
\label{ppa}\\
\mbox{Tr}[\Gamma_{\mu 5}\sigma_z\gamma_{\nu}]&=&
-4 \epsilon_{\mu\nu 12} \\
\mbox{Tr}[\Gamma_{\mu 5}\gamma_{\nu}\sigma_z]&=& - 4\epsilon_{\mu\nu 12}.
\label{ppb}
\end{eqnarray}
Given these equations, it would be easier to evaluate the Dirac traces
appearing in Eqn. [\ref{trace1}]. We would start with the piece proportional
 to $m^2$. It reads,
\begin{eqnarray}
m^2 \left( \Gamma_{\mu 5}\sigma_z\gamma_{\nu}T_s +
\Gamma_{\mu 5}\gamma_{\nu} \sigma_z T_s \right) = - 4 m^2\epsilon_{\mu\nu 12}
\left(T_s + T_s'\right).
\label{msq}
\end{eqnarray}
It is worth noting that, the sign of the piece proportional to $m^2$ is 
negative, more over the piece is antisymmetric with respect to the indices
$\mu$ and $\nu$. Next we would like to evaluate the trace of the pieces
having parallel component of momentums $p$ and $p'$, i.e.:  

\begin{eqnarray}
\mbox{Tr}\left[ \Gamma_{\mu 5}\left(
\sigma_z \rlap/p_{\para}\gamma_{\nu}\rlap/p'_{\para} T_s+
\rlap/p_{\para}\gamma_{\nu} \sigma_z\rlap/p'_{\para} T_s'
\right) \right] &=&  \mbox{Tr}\left[\gamma_{\mu}\gamma_{\alpha_{\para}}
\gamma_{\nu}\gamma_{\beta_{\para}}p^{'\beta_{\para}} 
p^{\widetilde{\alpha_{\para}}} T_s
 +\gamma_{\mu}
\gamma_{\beta_{\para}}
\gamma_{\nu}\gamma_{\alpha_{\para}}p^{\beta_{\para}} 
p^{'\widetilde{\alpha_{\para}}} T_s'\right] \nonumber \\
&=& 4\left[ 
\left( g_{\mu \alpha\para} p^{\widetilde{\alpha\para}}
p'_{\nu\para} - g_{\mu \nu} p'_{\alpha\para}
p^{\widetilde{\alpha\para}} +g_{\nu \alpha\para} p^{\widetilde{\alpha\para}}
p'_{\mu\para}\right)T_s \nonumber\right.\\ \left.
\right. &&+ \left.
\left(g_{\mu \alpha\para} p'^{\widetilde{\alpha\para}}
p_{\nu\para} - g_{\mu \nu} p_{\alpha\para}
p'^{\widetilde{\alpha\para}} +g_{\nu \alpha\para} 
p'^{\widetilde{\alpha\para}}
p_{\mu\para} \right)T_s'
\right].
\label{trpara}
\end{eqnarray}
It should be noted that unlike the momentum dependent piece, this part is 
symmetric in the indices $\mu$ and $\nu$. In other words these terms are 
 insensitive to the position of $\gamma_5$. That is to say,inside the trace, 
 if we would 
have pushed the $\gamma_5$ matrix to the $\mu$ vertex, these pieces would 
remain insensitive.  Its also important to notice that these pieces are
symmetric with respect to exchange of the variables
 $(p,s)\Longleftrightarrow (p',s')$. The last trace that needs to be 
evaluated is,
\begin{eqnarray}
\mbox{Tr}\left[\Gamma_{\mu 5}\sigma_z \rlap/p_{\para}
\gamma_{nu}\rlap/p'_{\perp}T_s S_{s'}
+ \Gamma_{\mu 5}\rlap/p_{\perp} \gamma_{nu}\sigma_z \rlap/p'_{\para} 
T_{s'}S_s \right] &=& 4\left[ \left( g_{\mu \alpha_{\para}}p^{\widetilde{
\alpha_{\para}}}
p'_{\nu_\perp}+ g_{\nu \alpha_{\para}}p^{\widetilde{\alpha_{\para}}}
p'_{\mu_\perp}
\right) T_s S_{s'} \right. \nonumber \\ \left.
\right. & + & \left.
\left( g_{\mu \alpha_{\para}}p^{'\widetilde{\alpha_{\para}}}
p_{\nu_\perp}+ g_{\nu \alpha_{\para}}p^{'\widetilde{\alpha_{\para} }}
p_{\mu_\perp}
\right) T_{s'} S_s \right]
\label{trprp}
\end{eqnarray}
As can be verified, Eqn. [\ref{trprp}] is symmetric is the indices 
$\mu \& \nu $, more over as before it remains the same under the 
transformation of variables $(p,s)\Longleftrightarrow (p',s')$.
On the other hand the symmetry under the exchange of
$(p,s)\Longleftrightarrow (p',s')$ can be exploited to evaluate the
second trace in Eqn.(\ref{complapp1}). Since the momentum dependent
terms all appear in even powers of $p$ and $p'$,these equations are 
insensitive to the following transformation 
$(p,p') \leftrightarrow (-p',-p)$. Hence under the transformation
\begin{eqnarray}  
\left\{p,s\right\} \Longleftrightarrow \left\{-p',s' \right\}
\label{pp'ss'symm},
\end{eqnarray}
the traces would remain invariant. Therefore the Dirac trace of both these
pieces are the same and hence upon using the 
definitions of, $T_s$, $T_{s'}$, $S_{s}$ and $S_{s'}$ as given in 
Eqn.[\ref{short}], we recover equation [\ref{roddm}].
%
%
\renewcommand{\thesection}{Appendix}
\renewcommand{\theequation}{\Alph{section}.\arabic{equation}}
\section*{appendix} 
%
\subsection*{Appnedix B: Proof Of $~~k^{\mu}R^{(0)}_{\mu\nu_{\perp}} = 0 $}


The expression for $\mbox{R}^{(o)}_{\mu \nu_{\perp}}(k)$ 
as follows from Eqn. [\ref{rodd-1}], turns out to be,
\begin{eqnarray}
\mbox{R}^{(o)}_{\mu\nu_{\perp}}=4i\eta_{+}(p)\Bigg[ 
\Bigg\{ - g_{\mu \nu_{\perp}} k_{\alpha\para}p^{\widetilde{\alpha\para}} 
 + g_{\mu \alpha\para} p^{\widetilde{\alpha\para}}
p'_{\nu\pr} \sec^2(eQ_f{\cal B}s')\Bigg\} \tan(eQ_f{\cal
B}s)  \nonumber \\
+ \Bigg\{ - g_{\mu \nu_{\perp}} p_{\alpha\para}k^{\widetilde{\alpha\para}} 
+ g_{\mu \alpha\para} p'^{\widetilde{\alpha\para}}
p_{\nu\pr} \sec^2(eQ_f{\cal B}s')\Bigg\} \tan(eQ_f{\cal
B}s)
\Bigg]
\label{rodd-perp}
\end{eqnarray}

Contracting $\mbox{R}^{(o)}_{\mu\nu_{\perp}}$ with $k^{\mu}$, we get,

\begin{eqnarray}
k^{\mu}\mbox{R}^{(o)}_{\mu\nu_{\perp}}=4i\eta_{+}(p)\Bigg[ 
\Bigg\{ - k_{\nu_{\perp}} k_{\alpha\para}p^{\widetilde{\alpha\para}} 
 + k_{\alpha\para} p^{\widetilde{\alpha\para}}
p'_{\nu\pr} \sec^2(eQ_f{\cal B}s')\Bigg\} \tan(eQ_f{\cal
B}s)  \nonumber \\
+ \Bigg\{ - k_{ \nu_{\perp}} p_{\alpha\para}k^{\widetilde{\alpha\para}} 
+ k_{ \alpha\para} p^{\widetilde{\alpha\para}}
p_{\nu\pr} \sec^2(eQ_f{\cal B}s')\Bigg\} \tan(e Q_f{\cal
B}s
\Bigg]
\label{rodd-perp1}
\end{eqnarray}

In getting Eqn. [\ref{rodd-perp1}] the following definition was used,
\begin{eqnarray}
a_{\mu_{\para}} b^{{\widetilde \mu}\para}=a_0 b^3 + a_3 b^0.
\label{deftilde}
\end{eqnarray}
It can be easily verified that, if $a$ and $b$ are the same, 
Eqn.[\ref{deftilde}], vanishes. This fact was additionally exploited while 
arriving at, Eqn.[\ref{rodd-perp1}]. In particular, if we use relations,
[\ref{primeperpint}, \ref{pperpint}], for $p_{\nu\pr}$ $p'_{\nu\pr}$ as 
appearing in Eqn.[\ref{rodd-perp1}], the same simplifies further to

\begin{eqnarray}
k^{\mu}\mbox{R}^{(o)}_{\mu\nu_{\perp}}=&&\!\!\!\!\!4i\eta_{+}(p)\left[ 
 k_{\nu_{\perp}} k_{\alpha\para}p^{\widetilde{\alpha\para}} 
\left(\tan (eQ_f{\cal B}s')- \tan(eQ_f{\cal B}s)\right)+k_{\nu_{\perp}} 
k_{\alpha\para}
p^{\widetilde{\alpha\para}}\right. \nonumber \\ &\times& \left.
\frac{\left(\tan^2(eQ_f{\cal B}s)- \tan^2(eQ_f{\cal B}s')\right)}{\left(\tan(eQ_f{\cal B}s)+\tan(eQ_f{\cal B}s')\right)}
\right] = 0.
\label{rodd-perp2}
\end{eqnarray}
\noindent
So we have proved that, $k^{\mu}R^{(o)}_{\mu\nu_{\perp}}=0$. \\


\renewcommand{\thesection}{Appendix}
\renewcommand{\theequation}{\Alph{section}.\arabic{equation}}
\section*{appendix} 

\subsection*{Appendix C: Proof Of $ k^{\mu}R^{(0.1)}_{\mu\nu_{\para}} = 0 $}
%
In the text, we claimed that the contribution coming from Eqn. [\ref{pal1}]
must vanish. Here, we justify that claim. 
Vanishing of Eqn.[{\ref{pal1}}], actually amounts to showing
that the following integral,
\begin{eqnarray}
C_{1\widetilde{\nu_{\para}}} &=& \int \frac{d^4p}{(2\pi)^4}~ \eta_+(p)
\int_{-\infty}^\infty\!\!\! ds \!\!\! e^{\Phi(p,s)}
\int_0^\infty \!\!\! ds'\!\! e^{\Phi(p',s')} 
 p_{\widetilde{\nu_{\para}}}\!\! \Bigg[
\left( k^2_{\para} +2 (k \cdot p)_{\para}\!\!\right)\left( \tan(eQ_f{\cal B}s) +\tan(eQ_f{\cal B}s')
\right)   \nonumber  \\
&-& k^2_{\perp} (\tan(eQ_f{\cal B}s)- \tan(eQ_f{\cal B}s'))\Bigg] = 0.
\label{pal2}
\end{eqnarray}
%

We begin by defining two new parameters
\begin{eqnarray}
\xi &=& \frac12 eQ_f{\cal B}(s+s') \,, \nonumber\\*
\zeta &=& \frac12 e Q_f{\cal B}(s-s') \,.
\label{xizeta}
\end{eqnarray}
and noting, ( following \cite{frd1} ), that:
\begin{eqnarray}
{ie{\cal B}} \; {d\over d\zeta} e^{\Phi(p,s) + \Phi(p',s')} = 
e^{\Phi(p,s) + \Phi(p',s')} \left(
k_\parallel^{2} + 2\left(p.k \right)_\parallel - p_\perp^{\prime2} \sec^2
(\xi-\zeta) + 
p_\perp^2 \sec^2 (\xi+\zeta) \right) \,.
\label{C1par}
\end{eqnarray}
%

Using Eqns. [\ref{psq}] and [\ref{p'sq}]  in Eqn. [\ref{C1par}] and little bit 
of algebra, one can further show that, 
%
\begin{eqnarray}
C_{1\widetilde{\nu_{\para}}}=
ieQ_f{\cal B} \int \frac{d^4p}{(2\pi)^4} p_{\widetilde{\nu}
_{\para}}\eta_+(p)
\int_{-\infty}^\infty \!\!\! ds \int_0^\infty ds' \; 
{d\over d\zeta} {\cal F}(\xi,\zeta) \,,
\end{eqnarray}
where
\begin{eqnarray}
{\cal F}(\xi,\zeta) =
\Big(\tan eQ_f{\cal B}s + \tan eQ_f{\cal B}s' \Big) 
e^{\Phi(p,s) + \Phi(p',s')}
\end{eqnarray}
with $s$ and $s'$ related to $\xi$ and $\zeta$ through Eq.\
(\ref{xizeta}). We can now change the integration variables to 
$\xi$ and $\zeta$ to arrive at (ignoring the unimportant constants
in front),
\begin{eqnarray}
C_{1\tilde{\nu_{\para}}} & \propto &  \int \frac{d^4p}{(2\pi)^4} 
p_{\widetilde{\nu}_{\para}}
\eta_+(p)
\int_{-\infty}^\infty d\xi \int_{-\infty}^\infty d\zeta \;
\Theta(\xi-\zeta) \; {d\over d\zeta} 
{\cal F}(\xi,\zeta) \nonumber\\*
&=&  \int \frac{d^4p}{(2\pi)^4} p_{{\tilde{\nu}_{\para}}} \eta_+(p)
\int_{-\infty}^\infty d\xi \int_{-\infty}^\infty d\zeta \;
\Big[ {d\over d\zeta} \Big\{ \Theta(\xi-\zeta) \; 
{\cal F}(\xi,\zeta) \Big\} - \delta(\xi-\zeta) \; 
{\cal F}(\xi,\zeta) \Big] \nonumber\\*
&=& -\, \int \frac{d^4p}{(2\pi)^4}p_{\tilde{\nu}_{\para}} \eta_+(p)
\int_{-\infty}^\infty d\xi \;
{\cal F}(\xi,\xi) \,,
\end{eqnarray}
since the other term vanishes at the limits. In this integrand,
$\zeta=\xi$, which means $s'=0$. Looking back at the definition of
$\cal F$, we find
\begin{eqnarray}
{\cal F}(\xi,\xi) = \exp \left\{\Phi(p,{2\xi\over e Q_f{\cal B}})\right\}
\tan 2\xi \,. 
\end{eqnarray}
This is an even function of $p$, whereas 
$\left( p_{\widetilde{\nu}_{\para}}\!\!\! \times \eta_+(p) \right) $ is odd. Thus, the
expression vanishes on integration over $p$.
%
%
\renewcommand{\thesection}{Appendix}
\renewcommand{\theequation}{\Alph{section}.\arabic{equation}}
\section*{appendix} 

\subsection*{Appendix D: Proof Of $ k^{\mu}R^{(0.2)}_{\mu\nu_{\para}} = 0 $}
%
%
%
In this appendix we show that, $k^{\mu} R^{(0.2)}_{\mu\nu_{\para}} =0$. The two 
components  of $\nu_{\para}$ are 0 and 3. Respective components those 
need evaluation, are,
\begin{eqnarray}
k^{\mu}R^{(o.2)}_{\mu 0}= - 4i\eta_{+}(p)k^3\left[\left(p^2_{\para}-m^2 \right)
\left( \tan(eQ_f{\cal B}s)\!\!\! \right.\right.\!\ &+&\left.\left.\!\! \tan(eQ_f{\cal B}s')\right) - 
k^2_{\perp} \tan^2(eQ_f{\cal B}s') \right.\nonumber \\ &\times& \left.
\frac{\sec^2eQ_f{\cal B}s}{\tan(eQ_f{\cal B}s)
+ \tan(eQ_f{\cal B}s')}\right] \label{vertex0}
\end{eqnarray}
Similarly,
\begin{eqnarray}
 k^{\mu}R^{(o.2)}_{\mu 3}  = 4i\eta_{+}(p)k_0\left[\left(p^2_{\para}-m^2 
\right)
\left( \tan(eQ_f{\cal B}s)\!\!\! \right.\right. &+& \left. \left.
 \tan(eQ_f{\cal B}s')\right) - k^2_{\perp} \tan^2(eQ_f{\cal B}s')
\right.\nonumber \\ &\times& \left.
\frac{\sec^2eQ_f{\cal B}s}{\tan(eQ_f{\cal B}s)+ \tan(eQ_f{\cal B}s')}\right] 
\label{vertex3a}
\end{eqnarray}

 Expressions inside the square bracket of Eqns. [\ref{vertex0}] or 
[\ref{vertex3a}] are the same, 
introducing compact notation,
\begin{eqnarray}
\Pi(p,k)= 4i\eta_{+}(p)\left[\left(p^2_{\para}-m^2 \right)
\left( \tan(eQ_f{\cal B}s)+ \tan(eQ_f{\cal B}s')\right)\right.& -& \left. k^2_{\perp} \tan^2(eQ_f{\cal B}s') \right. \nonumber \\ &\times& \left. 
\frac{\sec^2eQ_f{\cal B}s}{\tan(eQ_f{\cal B}s)+ \tan(eQ_f{\cal B}s')}\right] 
\label{vertex3}
\label{Pi}
\end{eqnarray}
Actually the expression defined by $\Pi(p,k)$, appears inside the 
integral over the loop momentum variable $p$ as well as the parameters
$s$ and $s'$. Calling the integrals over $p$, $s$ and $s'$ of $\Pi(p,k)$
 as  $\hat{\Pi}(p,k)$, the same is:
\begin{eqnarray}
\hat{\Pi}(p,k)=\!\!\!\int\frac{d^4p}{(2\pi)^4}&&\!\!\!
\!\!\!\!\int^{\infty}_{-\infty}\!\int^{\infty}_{0}\!\!\!ds ds'
e^{\Phi(p,s)}e^{\Phi(p',s')}
 4i\eta_{+}(p)\nonumber 
\left[\left(p^2_{\para}-m^2 \right)
\!\!\left(\tan(eQ_f{\cal B}s) + \tan(eQ_f{\cal B}s')\right) 
\right.\nonumber \\ \left.
\right.&-&\left. k^2_{\perp} \tan^2(eQ_f{\cal B}s')
\frac{\sec^2(eQ_f{\cal B}s)}{\tan(eQ_f{\cal B}s)+ \tan(eQ_f{\cal B}s')}\right] 
\label{Pihat}
\end{eqnarray}\\
Though not obvious from appearance, but we would show that Eqn. [\ref{Pihat}] 
vanishes identically. To show that, 
we note that it is possible to write,
\begin{eqnarray}
\int^{\infty}_{-\infty}\!\!\!\!\!\! ds 
e^{is(p^2_{\para} - m^2) -i\frac{\tan(eQ_f{\cal B}s)}{eQ_f{\cal B}}
p^2_{\perp} - \epsilon |s|}
(p^2_{\para}-m^2)
\!\!\!&& \!\!\!\!\!\left( \tan(eQ_f{\cal B}s)+ \tan(eQ_f{\cal B}s')\right) 
= \!\!
\int^{\infty}_{-\infty}\!\!\!\!\!\!\! ds\! \left\{\!\!
\frac{d}{dis}e^{is(p^2_{\para} - m^2)} \right\} \nonumber \\
&\times&
e^{ -i\frac{\tan(eQ_f{\cal B}s)}{eQ_f{\cal B}}p^2_{\perp} -\epsilon |s|} 
\left(\tan(eQ_f{\cal B}s)+\tan(eQ_f{\cal B}s') \right)
\label{Pihat1}
\end{eqnarray}
\\
Upon partial integration of Eqn. [\ref{Pihat1}],i.e.,
\begin{eqnarray}
\int^{\infty}_{-\infty}\!\!\!\!\!&ds&\!
\left\{\!\!\frac{d}{dis}e^{is(p^2_{\para} - m^2)} \!\right\}
\!\! 
\,\,e^{ -i\frac{\tan(eQ_f{\cal B}s)}{eQ_f{\cal B}}p^2_{\perp} -\epsilon |s|} 
\left(\tan(eQ_f{\cal B}s)+\tan(eQ_f{\cal B}s') \right)
\nonumber \\ &=& \!
-\!\!\int^{\infty}_{-\infty}\!\!\!\!\! \!\!\!\!ds 
e^{is(p^2_{\para} - m^2)}\!\!\! \frac{d}{dis}\left\{
e^{ -i\frac{\tan(eQ_f{\cal B}s)}{eQ_f{\cal B}}p^2_{\perp} -\epsilon |s|} 
\right\} 
\left(\tan(eQ_f{\cal B}s)+\tan(eQ_f{\cal B}s') \right)
 \nonumber \\
&&-\int^{\infty}_{-\infty} \!\!\!\!\!\!ds 
e^{is(p^2_{\para} - m^2)
 -i\frac{\tan(eQ_f{\cal B}s)}{eQ_f{\cal B}}p^2_{\perp} -\epsilon |s|}
\frac{d}{dis}\left(\tan(eQ_f{\cal B}s)+\tan(eQ_f{\cal B}s') \right)
\label{partint}
\end{eqnarray}
It should be noted that, while deriving eqn. [\ref{partint}]
a total derivative term in has been thrown away, since the integrand
vanishes at the boundary (virtue of the $\epsilon$ prescription). 
Of the terms on the right hand side of Eqn. [\ref{partint}], 
\begin{eqnarray}
\!\!\int^{\infty}_{-\infty} \!\!\!\!\!\!\!ds 
e^{is(p^2_{\para} - m^2)}&&\!\!\!\!\!\!\!\! 
\frac{d}{dis}\!\!\left\{
e^{ -i\frac{\tan(eQ_f{\cal B}s)}{eQ_f{\cal B}}p^2_{\perp} -\epsilon |s|} 
\right\}
\left(\tan(eQ_f{\cal B}s) + \tan(eQ_f{\cal B}s') \right) \nonumber \\
&=& (-1)
\!\!\int^{\infty}_{-\infty} \!\!\!\!ds 
e^{\Phi(p,s)}  p^2_{\perp} sec^2(e {\cal B}) 
\left(\tan(eQ_f{\cal B}s)+\tan(eQ_f{\cal B}s') \right)
\label{xxa}
\end{eqnarray}
and,
\begin{eqnarray}
\int^{\infty}_{-\infty} \!\!\!\!\!\!\!ds  
e^{is(p^2_{\para} - m^2)
 -i\frac{\tan(eQ_f{\cal B}s)}{eQ_f{\cal B}}p^2_{\perp} -\epsilon |s|} && 
\!\!\!\!\!\!\!
\frac{d}{dis}\left(\tan(eQ_f{\cal B}s)+\tan(eQ_f{\cal B}s') \right)
\nonumber \\
&=&-ieQ_f{\cal B} \int^{\infty}_{-\infty} e^{\Phi(p,s)}
sec^2(eQ_f{\cal B}s)ds. 
\label{xxb}
\end{eqnarray}
\\
The expression for $\Phi(p,s)$ is the same as provided earlier. Therefore
using Eqns.[ \ref{xxa}] and [\ref{xxb} ]
\begin{eqnarray}
\int^{\infty}_{-\infty}\!\!\! \!\!ds &&\!\!\!\!
e^{\Phi(p,s)}
(p^2_{\para} - m^2)\,\,
\!\!\!\left(\tan(eQ_f{\cal B}s)+ \tan(eQ_f{\cal B}s')\right) 
=
\int^{\infty}_{-\infty}\!\!\!\! ds 
e^{\Phi} \left[ p^2_{\perp} \sec^{2}(eQ_f{\cal B})
\left(\tan(eQ_f{\cal B}s)
\right.\right. \nonumber \\
\!\!\!&+& \left.\left.
\!\!\!\tan(eQ_f{\cal B}s') \right)
+
 \!  ieQ_f{\cal B}
\sec^{2}(eQ_f{\cal B}) \right] 
=\!\!\!\! \int^{\infty}_{-\infty}\!\!\!\!\!ds e^{\Phi(p,s)}\!\!\left[ 
\frac{\tan^2(eQ_f{\cal B}s')sec^2(eQ_f{\cal B}s)}{\left(\tan(eQ_f{\cal B}s)+
\tan(eQ_f{\cal B}s' )\right)} k^{2}_{\perp}
\right]
\label{totderi2}
\end{eqnarray}
\\
Where in the last step of Eqn. [\ref{totderi2}], one has used the identity
given by Eqn. [\ref{psq}]. It can now be easily verified, that on substituting
Eqn.[\ref{totderi2}] in Eqn.[\ref{Pihat}] the pieces proportional to 
$k^2_{\perp}$ compensate each other. So the proof that, 
$ k^{\mu}R^{(0.2)}_{\mu\nu_{\para}} = 0 $ is complete.

%
%
%


%

\end{document}